\pdfoutput=1
\documentclass[12pt,a4paper]{article}

\usepackage{ifthen} 
\newboolean{pdflatex}
\setboolean{pdflatex}{true}

\newboolean{articletitles}
\setboolean{articletitles}{true} 
\newboolean{uprightparticles}
\setboolean{uprightparticles}{false} 

\def\paperauthors{LHCb collaboration} 
\def\paperasciititle{Search for resonances decaying to photon pairs with masses between 4.9 and 19.4 GeV} 
\def\papertitle{Search for resonances decaying to photon pairs with masses between \mbox{4.9 and 19.4\,GeV}} 
\def\paperkeywords{{High Energy Physics}, {LHCb}, {exotics}, {branching fraction}, {particle and resonance production}, {calorimeter}, {ALP}} 

\def\papercopyright{\the\year\ CERN for the benefit of the LHCb collaboration} 
\def\paperlicence{CC BY 4.0 licence}
\def\paperlicenceurl{https://creativecommons.org/licenses/by/4.0/}

\newif\ifEnableSectionTOCLinks

\EnableSectionTOCLinksfalse %

\usepackage[top=1in, bottom=1.25in, left=1in, right=1in]{geometry}

\usepackage{microtype}
\usepackage{lineno}  
\usepackage{xspace} 
\usepackage{caption} 

\usepackage{graphicx}  
\usepackage{color}
\usepackage{colortbl}
\graphicspath{{./figs/}} 

\usepackage{amsmath} 
\usepackage{amssymb}
\usepackage{amsfonts}
\usepackage{upgreek} 

\newcommand*\patchAmsMathEnvironmentForLineno[1]{%
\expandafter\let\csname old#1\expandafter\endcsname\csname #1\endcsname
\expandafter\let\csname oldend#1\expandafter\endcsname\csname
end#1\endcsname
 \renewenvironment{#1}%
   {\linenomath\csname old#1\endcsname}%
   {\csname oldend#1\endcsname\endlinenomath}%
}
\newcommand*\patchBothAmsMathEnvironmentsForLineno[1]{%
  \patchAmsMathEnvironmentForLineno{#1}%
  \patchAmsMathEnvironmentForLineno{#1*}%
}
\AtBeginDocument{%
\patchBothAmsMathEnvironmentsForLineno{equation}%
\patchBothAmsMathEnvironmentsForLineno{align}%
\patchBothAmsMathEnvironmentsForLineno{flalign}%
\patchBothAmsMathEnvironmentsForLineno{alignat}%
\patchBothAmsMathEnvironmentsForLineno{gather}%
\patchBothAmsMathEnvironmentsForLineno{multline}%
\patchBothAmsMathEnvironmentsForLineno{eqnarray}%
}

\usepackage[pdftex,
            pdfauthor={\paperauthors},
            pdftitle={\paperasciititle},
            pdfkeywords={\paperkeywords}]{hyperref}

\usepackage{hyperxmp}

\usepackage[colorinlistoftodos,textsize=scriptsize]{todonotes}

\usepackage[bottom,flushmargin,hang,multiple]{footmisc}

\usepackage[all]{hypcap} 

\usepackage{xspace} 
\usepackage{upgreek}

\def\lhcb   {\mbox{LHCb}\xspace}

\def\belle  {\mbox{Belle}\xspace}

\def\MagUp {\mbox{\em Mag\kern -0.05em Up}\xspace}

\ifthenelse{\boolean{uprightparticles}}%
{
 
 \def\Pgamma      {\ensuremath{\upgamma}\xspace}

 \def\Pmu         {\ensuremath{\upmu}\xspace}

 \def\Ppi         {\ensuremath{\uppi}\xspace}

 \def\Ppsi        {\ensuremath{\uppsi}\xspace}

 \def\PDelta      {\ensuremath{\Delta}\xspace}                 
 \def\PXi         {\ensuremath{\Xi}\xspace}                 
 \def\PLambda     {\ensuremath{\Lambda}\xspace}                 
 \def\PSigma      {\ensuremath{\Sigma}\xspace}                 
 \def\POmega      {\ensuremath{\Omega}\xspace}                 
 \def\PUpsilon    {\ensuremath{\Upsilon}\xspace}
 \let\oldPi\Pi
 \def\PPi         {\ensuremath{\oldPi}\xspace}

 \def\PB      {\ensuremath{\mathrm{B}}\xspace}                 
 \def\PD      {\ensuremath{\mathrm{D}}\xspace}                 
 \def\PJ      {\ensuremath{\mathrm{J}}\xspace}                 
 \def\PK      {\ensuremath{\mathrm{K}}\xspace}                 
 \def\Pb      {\ensuremath{\mathrm{b}}\xspace}

 \def\Ps      {\ensuremath{\mathrm{s}}\xspace}

 \def\thebaroffset{0.0em}
}
{
 
 \def\Pgamma      {\ensuremath{\gamma}\xspace}

 \def\Pmu         {\ensuremath{\mu}\xspace}

 \def\Ppi         {\ensuremath{\pi}\xspace}

 \def\Ppsi        {\ensuremath{\psi}\xspace}                 
                  
 \mathchardef\PDelta="7101
 \mathchardef\PXi="7104
 \mathchardef\PLambda="7103
 \mathchardef\PSigma="7106
 \mathchardef\POmega="710A
 \mathchardef\PUpsilon="7107
 \mathchardef\PPi="7105
 \def\PB      {\ensuremath{B}\xspace}                 
 \def\PD      {\ensuremath{D}\xspace}                 
 \def\PJ      {\ensuremath{J}\xspace}                 
 \def\PK      {\ensuremath{K}\xspace}                 
 \def\Pb      {\ensuremath{b}\xspace}

 \def\Ps      {\ensuremath{s}\xspace}

 \def\thebaroffset{0.18em}
}
\newcommand{\offsetoverline}[2][\thebaroffset]{\kern #1\overline{\kern -#1 #2}}%

\makeatletter
\ifcase \@ptsize \relax
  \newcommand{\miniscule}{\@setfontsize\miniscule{4}{5}}
\or
  \newcommand{\miniscule}{\@setfontsize\miniscule{5}{6}}
\or
  \newcommand{\miniscule}{\@setfontsize\miniscule{5}{6}}
\fi
\makeatother

\DeclareRobustCommand{\optbar}[1]{\shortstack{{\miniscule (\rule[.5ex]{1.25em}{.18mm})}
  \\ [-.7ex] $#1$}}

\def\mup        {{\ensuremath{\Pmu^+}}\xspace}
\def\mun        {{\ensuremath{\Pmu^-}}\xspace} 

\def\g      {{\ensuremath{\Pgamma}}\xspace}

\def\squark    {{\ensuremath{\Ps}}\xspace}

\def\bquark    {{\ensuremath{\Pb}}\xspace}
\def\bquarkbar {{\ensuremath{\overline \bquark}}\xspace}
\def\bbbar     {{\ensuremath{\bquark\bquarkbar}}\xspace}

\def\pion   {{\ensuremath{\Ppi}}\xspace}
\def\piz    {{\ensuremath{\pion^0}}\xspace}

\def\pim    {{\ensuremath{\pion^-}}\xspace}

\def\kaon    {{\ensuremath{\PK}}\xspace}

\def\KorKbar {\kern \thebaroffset\optbar{\kern -\thebaroffset \PK}{}\xspace}

\def\Kp      {{\ensuremath{\kaon^+}}\xspace}

\def\Kstarz  {{\ensuremath{\kaon^{*0}}}\xspace}

\def\D       {{\ensuremath{\PD}}\xspace}

\def\DorDbar {\kern \thebaroffset\optbar{\kern -\thebaroffset \PD}\xspace}

\def\Dp      {{\ensuremath{\D^+}}\xspace}
\def\Dm      {{\ensuremath{\D^-}}\xspace}

\def\DpDm    {\ensuremath{\Dp {\kern -0.16em \Dm}}\xspace}

\def\B       {{\ensuremath{\PB}}\xspace}

\def\BorBbar {\kern \thebaroffset\optbar{\kern -\thebaroffset \PB}\xspace}

\def\Bd      {{\ensuremath{\B^0}}\xspace}

\def\BdorBdbar {\kern \thebaroffset\optbar{\kern -\thebaroffset \Bd}\xspace}

\def\Bs      {{\ensuremath{\B^0_\squark}}\xspace}

\def\BsorBsbar {\kern \thebaroffset\optbar{\kern -\thebaroffset \Bs}\xspace}

\def\Bds     {{\ensuremath{\B_{(\squark)}^0}}\xspace}

\def\jpsi     {{\ensuremath{{\PJ\mskip -3mu/\mskip -2mu\Ppsi}}}\xspace}

\def\Y#1S{\ensuremath{\PUpsilon{(#1S)}}\xspace}
\def\OneS  {{\Y1S}\xspace}

\def\LorLbar     {\kern \thebaroffset\optbar{\kern -\thebaroffset \PLambda}\xspace}

\newcommand{\decay}[2]{\mbox{\ensuremath{#1\!\to #2}}\xspace} 

\def\to                 {\ensuremath{\rightarrow}\xspace}

\def\AT#1     {\ensuremath{A_{\mathrm{T}}^{#1}}\xspace}           

\def\C#1      {\ensuremath{\mathcal{C}_{#1}}\xspace}                       
\def\Cp#1     {\ensuremath{\mathcal{C}_{#1}^{'}}\xspace}                    
\def\Ceff#1   {\ensuremath{\mathcal{C}_{#1}^{\mathrm{(eff)}}}\xspace}        
\def\Cpeff#1  {\ensuremath{\mathcal{C}_{#1}^{'\mathrm{(eff)}}}\xspace}       
\def\Ope#1    {\ensuremath{\mathcal{O}_{#1}}\xspace}                       
\def\Opep#1   {\ensuremath{\mathcal{O}_{#1}^{'}}\xspace}                    

\newcommand{\nospaceunit}[1]{\ensuremath{\text{#1}}}       
\newcommand{\aunit}[1]{\ensuremath{\text{\,#1}}}       

\newcommand{\tev}{\aunit{Te\kern -0.1em V}\xspace}
\newcommand{\gev}{\aunit{Ge\kern -0.1em V}\xspace}
\newcommand{\mev}{\aunit{Me\kern -0.1em V}\xspace}
\newcommand{\kev}{\aunit{ke\kern -0.1em V}\xspace}
\newcommand{\ev}{\aunit{e\kern -0.1em V}\xspace}
 
\newcommand{\mevc}{\ensuremath{\aunit{Me\kern -0.1em V\!/}c}\xspace}
\newcommand{\gevc}{\ensuremath{\aunit{Ge\kern -0.1em V\!/}c}\xspace}
\newcommand{\mevcc}{\ensuremath{\aunit{Me\kern -0.1em V\!/}c^2}\xspace}
\newcommand{\gevcc}{\ensuremath{\aunit{Ge\kern -0.1em V\!/}c^2}\xspace}

\def\mub{\ensuremath{\,\upmu\nospaceunit{b}}\xspace}

\def\pb {\aunit{pb}\xspace}

\def\fb   {\ensuremath{\aunit{fb}}\xspace}
\def\invfb   {\ensuremath{\fb^{-1}}\xspace}

\def\ps   {\ensuremath{\aunit{ps}}\xspace}

\newcommand{\chisq}{\ensuremath{\chi^2}\xspace}

\def\gsim{{~\raise.15em\hbox{$>$}\kern-.85em
          \lower.35em\hbox{$\sim$}~}\xspace}
\def\lsim{{~\raise.15em\hbox{$<$}\kern-.85em
          \lower.35em\hbox{$\sim$}~}\xspace}

\def\pt         {\ensuremath{p_{\mathrm{T}}}\xspace}

\def\pythia     {\mbox{\textsc{Pythia}}\xspace}

\def\tell1  {TELL1\xspace}
\def\ukl1   {UKL1\xspace}

\newcommand{\lhcborcid}[1]{\href{https://orcid.org/#1}{\hspace*{0.1em}\raisebox{-0.45ex}{\includegraphics[width=1em]{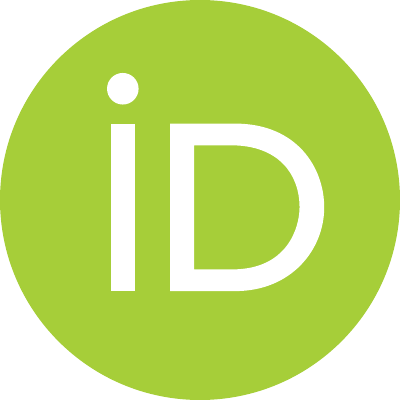}}}}

\hypersetup{
  colorlinks   = true, 
  urlcolor     = blue, 
  linkcolor    = blue, 
  citecolor    = red   
}

\ifEnableSectionTOCLinks
    \usepackage[explicit]{titlesec} 
    
    \let\oldcontentsline\contentsline
    \renewcommand

    \titleformat{\section}{\normalfont\Large\bf}{\hyperlink{tocsection.\thesection}{{\thesection} \parbox[t]{\dimexpr\textwidth-1pc}{#1}}}{1pc}{}

    \titleformat{\subsection}{\normalfont\bf}{\hyperlink{tocsubsection.\thesubsection}{{\thesubsection} \parbox[t]{\dimexpr\textwidth-1pc}{#1}}}{1pc}{}

    \titleformat{name=\section,numberless}[display]{}{}{0pt}{\normalfont\Huge\bfseries #1}
\fi

\usepackage{cite} 
\usepackage{mciteplus}

\def\madgraph{{\sc MadGraph5\_aMC@NLO} \xspace}
\def\etammg{\decay{\eta}{\mup \mun \g}}
\def\aa{{\ensuremath{\gamma\gamma}}\xspace}

\begin{document}

\renewcommand{\thefootnote}{\fnsymbol{footnote}}
\setcounter{footnote}{1}


\begin{titlepage}
\pagenumbering{roman}

\vspace*{-1.5cm}
\centerline{\large EUROPEAN ORGANIZATION FOR NUCLEAR RESEARCH (CERN)}
\vspace*{1.5cm}
\noindent
\begin{tabular*}{\linewidth}{lc@{\extracolsep{\fill}}r@{\extracolsep{0pt}}}
\ifthenelse{\boolean{pdflatex}}
{\vspace*{-1.5cm}\mbox{\!\!\!\includegraphics[width=.14\textwidth]{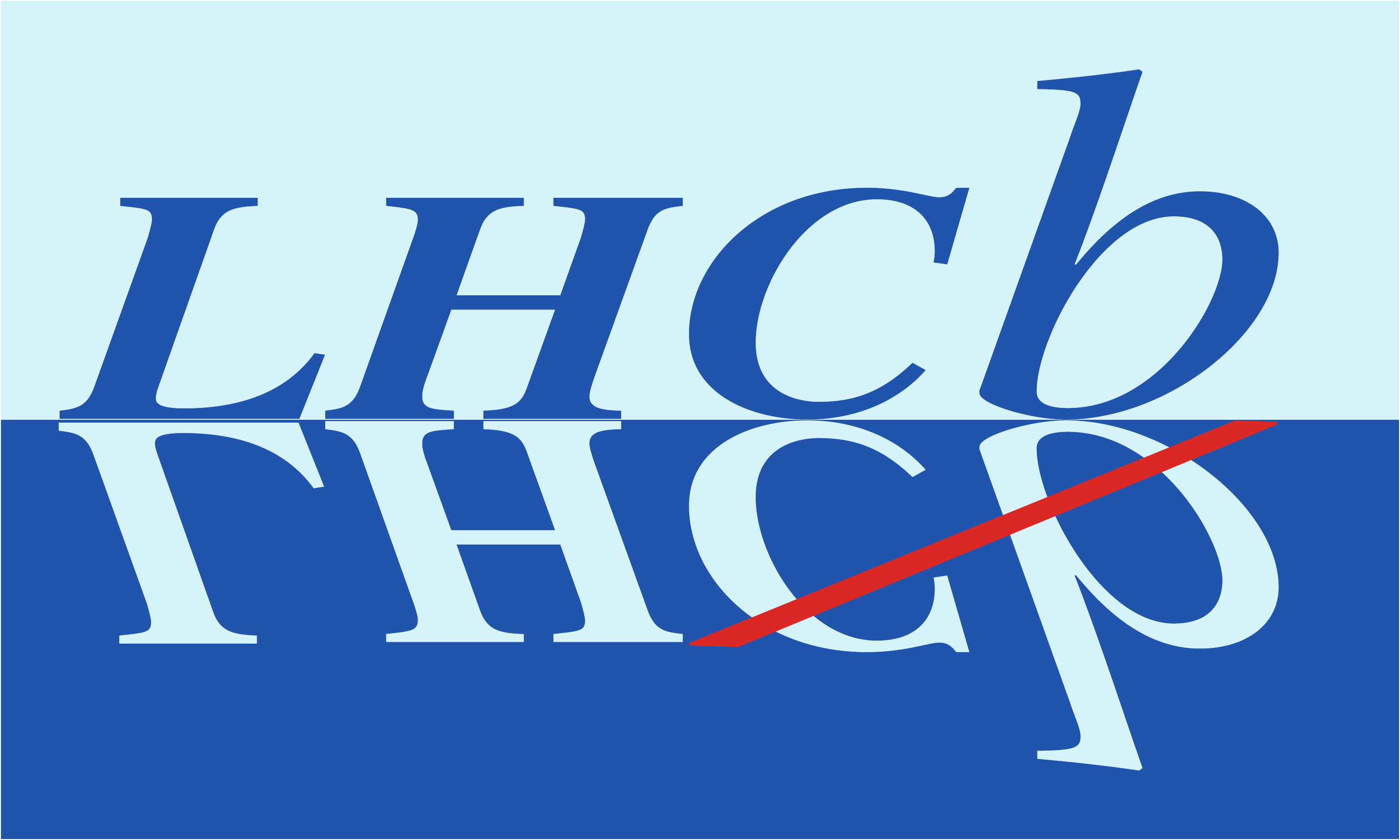}} & &}%
{\vspace*{-1.2cm}\mbox{\!\!\!\includegraphics[width=.12\textwidth]{figs/lhcb-logo.pdf}} & &}%
\\
 & & CERN-EP-2025-133 \\  
 & & LHCb-PAPER-2025-012 \\  
 & & July 18, 2025 \\ 
 & & \\
\end{tabular*}

\vspace*{4.0cm}

{\normalfont\bfseries\boldmath\huge
\begin{center}
  \papertitle 
\end{center}
}

\vspace*{2.0cm}

\begin{center}
\paperauthors\footnote{Authors are listed at the end of this paper.}
\end{center}

\vspace{\fill}

\begin{abstract}
  \noindent
 A search is presented for axion-like particles (ALPs) with masses between 4.9 and 19.4\gev decaying to a pair of photons, using proton-proton collisions collected with the LHCb detector during 2018 at a centre-of-mass energy of $13\tev$, corresponding to an integrated luminosity of 2.1\invfb. The same strategy and sample is used to search for the decays of the \Bs, \Bd and $\eta_b$ mesons into photon pairs.
 No significant excess is found. Upper limits on the photon-pair branching fraction times the cross-section of ALP production are determined as a function of the ALP mass. Limits on the branching fractions of the beauty states are determined to be $\mathcal{B}(B^0_s\to\gamma\gamma)<2.7\times10^{-5}$, $\mathcal{B}(B^0\to\gamma\gamma)<0.83\times10^{-5}$, and $
 \sigma(pp\to\eta_b X)\times\mathcal{B}(\eta_b\to\gamma\gamma)<765\,\text{pb}$ at $95\,\%$ confidence level. 
\end{abstract}

\vspace*{2.0cm}

\begin{center}
  Submitted to Phys.~Rev.~Lett.
\end{center}

\vspace{\fill}

{\footnotesize 
\centerline{\copyright~\papercopyright. \href{\paperlicenceurl}{\paperlicence}.}}
\vspace*{2mm}

\end{titlepage}


\newpage
\setcounter{page}{2}
\mbox{~}
%
%
%
%

\renewcommand{\thefootnote}{\arabic{footnote}}
\setcounter{footnote}{0}

\cleardoublepage


\pagestyle{plain} 
\setcounter{page}{1}
\pagenumbering{arabic}


New particles decaying into photon pairs appear in many theories for physics beyond the Standard Model (SM).
In particular, theories introducing axion-like particles (ALPs) have attracted interest \cite{Holdom:1982ex,Choi:1988sy,Holdom:1985vx,Dine:1986bg,Flynn:1987rs,Rubakov:1997vp,Choi:1998ep,Berezhiani:2000gh,Hook:2014cda,Fukuda:2015ana,Dimopoulos:2016lvn,Agrawal:2017ksf,Gaillard:2018xgk, Bishara:2017pfq, Goh:2008xz,Mardon:2009gw,Fan:2010is,Bellazzini:2017neg, Ferretti:2013kya,Belyaev:2015hgo,Ferretti:2016upr}.
In these theories, the ALPs arise as pseudo Nambu--Goldstone bosons of a symmetry that is broken at an energy scale $f_a$. The coupling of an ALP field $a$ to the gauge sector of the SM then occurs through the terms in the Lagrangian~\cite{Mariotti:2017vtv,CidVidal:2018blh}
\begin{align}
 \mathcal{L} \supset \frac{a}{4\pi f_a}\left[ \alpha_3 c_3 G \tilde{G} + \alpha_2 c_2 W \tilde{W} + \alpha_1 c_1 B\tilde{B}\right] + \frac{1}{2} m^2_a a^2,
\label{eq:lagranagian}
\end{align}
where $m_a$ is the mass of the ALP field, and $G$, $W$ and $B$ are respectively the QCD, weak, and hypercharge fields with their Hermitian conjugate fields $\tilde{G}$, $\tilde{W}$ and $\tilde{B}$. The coupling constants $\alpha_3$, $\alpha_2$ and $\alpha_1$ set the strength of the strong and electroweak \mbox{$SU(3)_C \times SU(2)_L \times U(1)_Y$} interactions of the SM. The constants $c_i$ are anomaly coefficients which encode the coupling strength of the ALP to the SM, a standard candle being $c_1=c_2=c_3=10$  \cite{Mariotti:2017vtv,CidVidal:2018blh}. 

Concrete examples of theories beyond the SM containing this type of process are those including a heavy QCD axion~\cite{Holdom:1982ex,Choi:1988sy,Holdom:1985vx,Dine:1986bg,Flynn:1987rs,Rubakov:1997vp,Choi:1998ep,Berezhiani:2000gh,Hook:2014cda,Fukuda:2015ana,Dimopoulos:2016lvn,Agrawal:2017ksf,Gaillard:2018xgk}, ALP-mediated dark matter~\cite{CidVidal:2018blh,Bishara:2017pfq}, supersymmetry with an R-axion~\cite{Goh:2008xz,Mardon:2009gw,Fan:2010is,Bellazzini:2017neg}, and composite Higgs models~\cite{Ferretti:2013kya,Belyaev:2015hgo,Ferretti:2016upr}.
Unless \mbox{$c_3\ll c_1,c_2$}, the coupling to gluons naturally dominates that to photons and motivates the search for ALPs at hadron colliders.
The choice of a photon-pair final state over a hadronic final state provides a clean signature that can be more easily distinguished from backgrounds. 
Although several searches sensitive to these types of processes have been carried out at the {Ge\kern -0.1em V} mass scale~\cite{ATLAS:2020hii,ATLAS:2022abz,
CMS:2018erd,Dolan:2017osp,CHARM:1985anb,Riordan:1987aw,Blumlein:1990ay,NA64:2020qwq,Belle-II:2020jti,BESIII:2022rzz,BESIII:2024hdv,Halliday:2024lca,FASER:2024bbl,Jaeckel:2015jla,Knapen:2016moh,Capozzi:2023ffu,NOMAD:2000usb,PrimEx:2010fvg,Aloni:2019ruo,
Pybus:2023yex,CMS-PAS-HIG-24-014,Belle-II:2025bkt},\footnote{Natural units are used throughout this Letter.} ALP candidates with masses between 2\gev and 10\gev remain largely unconstrained. Indirect constraints on ALPs in this mass region could so far only be set using nonresonant $\gamma\gamma$
searches conducted at much higher masses~\cite{Gavela:2019cmq}. 
A phenomenological study~\cite{CidVidal:2018blh} has indicated that searching for \mbox{$a\to\gamma\gamma$} decays with the LHCb experiment can provide stringent direct limits in the so-far unexplored region between 5--10\gev and motivates the present search. 

The photon-pair final state also provides complementary sensitivity to physics beyond the SM through searches for $B^0_{(s)}\to\gamma\gamma$ decays. The rates of these decays are enhanced with respect to the SM in theories including
an extended Higgs sector~\cite{Aliev:1998va} or supersymmetry~\cite{Gemintern:2004bw}. Searches for these decays have been performed at the $B$ factories~\cite{BaBar:2010qny, Belle:2014sac, Belle:2024bsw}. The most stringent limits have been set by the \belle collaboration and correspond to \mbox{$\mathcal{B}(\decay{\Bd}{\gamma\gamma})<6.4\times10^{-8}$} and \mbox{$\mathcal{B}(B^0_s\to\gamma\gamma)<3.1\times10^{-6}\ \text{at $90\,\%$ CL},$} approaching the SM predictions \mbox{$\mathcal{B}(B^0\to\gamma\gamma)=\left(1.35^{+1.24}_{-0.75}\right)\times10^{-8}$} and \mbox{$\mathcal{B}(B^0_s\to\gamma\gamma)=\left(3.0^{+1.8}_{-1.6}\right)\times10^{-7}$}~\cite{Bosch:2002bv, Shen:2020hfq,Li:2022mwo}.

Furthermore, an investigation of resonances decaying to photon pairs with masses in the {Ge\kern -0.1em V} range can shed light on the production and decay of the little studied \bbbar bound state $\eta_b$.
This resonance, which has been seen in decays of higher-mass \bbbar resonances, has been measured to have a mass of \mbox{$9398.7\pm2.0
\mev$}~\cite{PDG2024} and has an estimated production cross-section of \mbox{$\sigma(pp\to\eta_bX)=2.7\mub$}
in the LHCb acceptance~\cite{Lansberg:2020ejc}. The branching fraction of \mbox{$\eta_b\to\gamma\gamma$} decays is predicted to be \mbox{$\mathcal{B}(\eta_b\to\gamma\gamma)=(4-5)\times10^{-5}$}~\cite{Lansberg:2020ejc, Colquhoun:2024wsj}. Searches for the similar process via photoproduction of the $\eta_b$ meson have been carried out at LEP~II without finding evidence~\cite{ALEPH:2002xdz,Abdallah:2006yg}.

This Letter presents a search for resonances decaying to photon pairs in the mass range \mbox{$[4.9,19.4]\gev$} with the LHCb experiment in proton-proton ($pp$) collisions collected at a centre-of-mass energy of 13\tev. The analysis is optimised for ALP and $\Bds$ decays, while using a similar selection strategy to search for \mbox{$\eta_b\to\gamma\gamma$} decays.
The data used were collected in 2018 and correspond to a total integrated luminosity of \mbox{$2.07\pm0.05\,\invfb$}.

The \lhcb detector is a single-arm forward
spectrometer covering the \mbox{pseudorapidity} range \mbox{$2<\eta <5$},
described in detail in Refs.~\cite{LHCb-DP-2008-001,LHCb-DP-2014-002}. The detector element most relevant to this analysis is the calorimeter system, which enables photons to be identified and their energies measured. This system consists of a scintillating-pad detector (SPD), a preshower detector, and electromagnetic (ECAL) and hadronic (HCAL) calorimeters~\cite{LHCb-DP-2020-001}.

Simulation is used to determine the signal efficiencies, mass models and background composition. Simulated events are produced with the software described in Refs.~\cite{Sjostrand:2007gs,Lange:2001uf,Agostinelli:2002hh,Allison:2006ve,LHCb-PROC-2011-006}.
In particular, the simulation of ALP production with different mass hypotheses uses~\madgraph~\cite{madgraph} with the NNPDF2.3LO PDF set~\cite{NNPDF:2014otw}, modelling parton showers with \pythia~\cite{Sjostrand:2007gs}.
The production of $\eta_b$ mesons is simulated following the model of $\OneS$ mesons in \pythia.

Candidate ALP decays are selected with an online and offline event selection.
The online event selection is performed by a trigger, which consists of a hardware stage and a two-level software stage.
At the hardware stage, a preliminary calorimeter clustering is performed. Events are required to have a photon cluster with \mbox{$\pt \gtrsim 3$\gev}, or an electron cluster with \mbox{$\pt \gtrsim 2.5$\gev}, associated with one of the signal photon candidates, where \pt is the momentum component transverse to the beam direction. Furthermore, busy events are discarded based on their hit multiplicity in the SPD.
At the software stage, a dedicated selection algorithm is used for this analysis, which is discussed in detail in Ref.~\cite{rta-pairs}.
At the first level, preliminary photon-pair candidates are constructed by combining the calorimeter clusters found at the hardware stage. Kinematic requirements are imposed on the momenta of these preliminary photon and photon-pair candidates, as well as mass requirements on the photon-pair system.
Two selections are used with different kinematic requirements, one targeting \mbox{$\Bds\to\gamma\gamma$} decays and one targeting higher mass ALP and $\eta_b$ resonances. The second software level performs a full calorimeter clustering to form photon and photon-pair candidates and applies, along with kinematic requirements, a multilayer perceptron classifier that combines photon identification information and the \pt of the photon-pair candidate to select events consistent with containing signal photon-pair decays.

In the offline selection, signal photon-pair candidates are constructed from two ECAL clusters that are assumed to originate from photons produced in the $pp$ collision.
This assumption holds for ALP lifetimes up to about $40\ps$. For larger lifetimes the mass resolution shows significant degradation.
Additional kinematic and photon identification requirements are imposed. Photons are required to have \mbox{$\pt>3\gev$} and \mbox{$p$ $>6\gev$}. The energy of the associated shower in the HCAL is allowed to be at most $10\,\%$ of that in the ECAL, and signal photons are not allowed to saturate the range of a calorimeter cell to ensure reliable energy measurements.
Multivariate methods are used to remove contributions from misidentified charged and neutral hadrons~\cite{LHCb-DP-2020-001}. 
Photon-pair candidates are required to have a mass in the range \mbox{$[4.9,19.4]\gev$}, while a sample corresponding to a tenth of the luminosity with masses below $4.8\gev$ is also retained and used to interpolate the background shape at the lower-mass boundary. 
The final selection step is achieved through a boosted decision tree classifier (BDT) with gradient boosting~\cite{xgboost}. The classifier is trained using a cocktail of simulated \mbox{$a\to\gamma\gamma$} decays as a signal proxy, with ALP masses at $6, 7, 9, 11, 13, 15, 17,$ and $19\gev$ in equal proportions, and $4\,\%$ of the data as background proxy.
The features included in the classifier exploit the fact that ALPs are predicted to be produced in isolation from other particles in the event, in contrast to the majority of background candidates.
They comprise the multiplicity and summed momenta of charged particles in a cone in the \mbox{$[\eta,\varphi]$}-plane around the momentum of the signal photon-pair candidate, where $\varphi$ is the azimuthal angle, and their relations to the signal photon-pair candidate.
A separate dedicated classifier is trained for the \mbox{$\Bds\to\gamma\gamma$} decay, adding the \Bds \pt as an input feature.

The branching fraction of \mbox{$\Bds\to\gamma\gamma$} decays is measured relative to \mbox{$\Bs\to\phi(1020) \gamma$} decay, where $\phi(1020) \to K^+ K^-$ and the $\phi(1020)$ meson is denoted as $\phi$ in the remainder of this Letter. Candidate \mbox{$\Bs\to\phi \gamma$} decays are selected in the same way as signal photon-pair candidates at the hardware stage of the trigger. The first-level software trigger requires at least one charged
particle with high \pt and inconsistent with originating from any proton-proton collision vertex (PV). For the second software trigger level and the subsequent offline selection, candidate $\phi$ decays are built from tracks loosely identified as kaons with a significant momentum and transverse momentum, forming a vertex displaced from the associated PV and having a combined mass within 15\mev of the known $\phi$ mass~\cite{PDG2024}. These candidates are combined with calorimeter clusters loosely identified as photons with $\pt>2.5\gev$ to build a $B$ meson candidate, which is required to have a mass close to its known mass~\cite{PDG2024}. The final selection step uses a BDT with gradient boosting trained using \mbox{$\Bs\to\phi \gamma$} simulation as a signal proxy and the $B$ mass sidebands as background proxy. The input feature set comprises topological features of the displaced two-hadron system.

For ALP candidates, a measurement of the production cross-section times the decay branching fraction is obtained via the formula
\begin{equation}
\sigma(pp\to a X)\times\mathcal{B}(a\to\gamma\gamma)=\frac{N(a\to\gamma\gamma)}{\varepsilon(a\to\gamma\gamma)\mathcal{L}},
\label{eq:BF}
\end{equation}
where $N$ denotes the yield, $\varepsilon$ the product of reconstruction and selection efficiencies, $\mathcal{L}$ the recorded luminosity and $X$ any additional particles in the decay.
The diphoton mass is fitted to directly determine \mbox{$\sigma(pp\to a X)\times\mathcal{B}(a \to\gamma\gamma)$}, accounting for the efficiency and the luminosity. An analogous formula is used for the $\eta_b$ case.
For the corresponding measurement with \Bds decays, a normalisation is performed relative to the \mbox{$\Bs\to\phi \gamma$} decay using the formula
\begin{align}
\mathcal{B}(\Bds\to\gamma\gamma)
&=
\frac{N(\Bds\to\gamma\gamma)\,\varepsilon(\Bs\to\phi\gamma)}
     {N(\Bs\to\phi\gamma)\,\varepsilon(\Bds\to\gamma\gamma)}
\notag\\
&\quad\times
\frac{f_{s}}{f_{(d,s)}}
\;\mathcal{B}\bigl(\Bs\to\phi(K^+K^-)\gamma\bigr).
\label{eq:BF_Ns}
\end{align}
allowing cancellation of the production cross-section and luminosity factors.
Here, the branching fraction \mbox{$\mathcal{B}(\Bs \to \phi(K^+K^-) \gamma)$} is taken to be \mbox{$(1.67\pm0.20)\times10^{-5}$}~\cite{PDG2024} and the fragmentation fraction ratio $\frac{f_s}{f_d}=0.2539\pm0.0079$~\cite{LHCb-PAPER-2020-046}.

The efficiencies for reconstructing and selecting the signal candidates are calculated using simulated events calibrated with data using photons from \etammg, \mbox{$\Bd \to\Kstarz(\Kp\pim) \gamma$}, and \mbox{$\Bs \to \phi \gamma$} decays as control modes.
In particular, \mbox{$\Bd \to \Kstarz \gamma$} and \mbox{$\Bs \to \phi \gamma$} decays are used to calibrate event multiplicity, the BDT selection efficiency and the mass resolution, while \etammg decays are used to calibrate trigger efficiency, calorimeter saturation, and particle identification variables exploiting the abundance of these decays in high-energy $pp$ collisions.
The trigger efficiencies are calibrated in bins of photon transverse momentum and pseudorapidity, while the photon identification efficiencies and the efficiency to not saturate a calorimeter cell range are calibrated in bins of the \chisq variable describing the spatial compatibility between the photon cluster and the closest track. Correlations between the two final-state photons are found to be negligible.
The event multiplicity is calibrated by determining a scaling factor between data and simulation in the control modes.

Systematic uncertainties are estimated for the calibration methods at each corresponding step.
In most cases, the calibration procedure is performed on simulation for both signal and control modes and evaluated by comparing efficiencies with and without the calibration.
The dominant uncertainties calculated in this way reach up to $12\,\%$ for the calibration of the particle identification efficiencies, driven by the limited amount of simulated \etammg decays with high photon \pt. 
A systematic uncertainty for the calibration of the event multiplicity is estimated from the variation in the data-simulation scale factor between the two $B$-meson control modes and reaches up to $5\,\%$.
Residual data-simulation differences in the $B$-meson kinematics are assessed with \mbox{$\Bs\to\jpsi(\mu^+\mu^-)\phi(K^+K^-)$} decays and a systematic uncertainty of $13\,\%$ is assigned.

\begin{figure}
  \centering
  \includegraphics[width=0.49\textwidth]{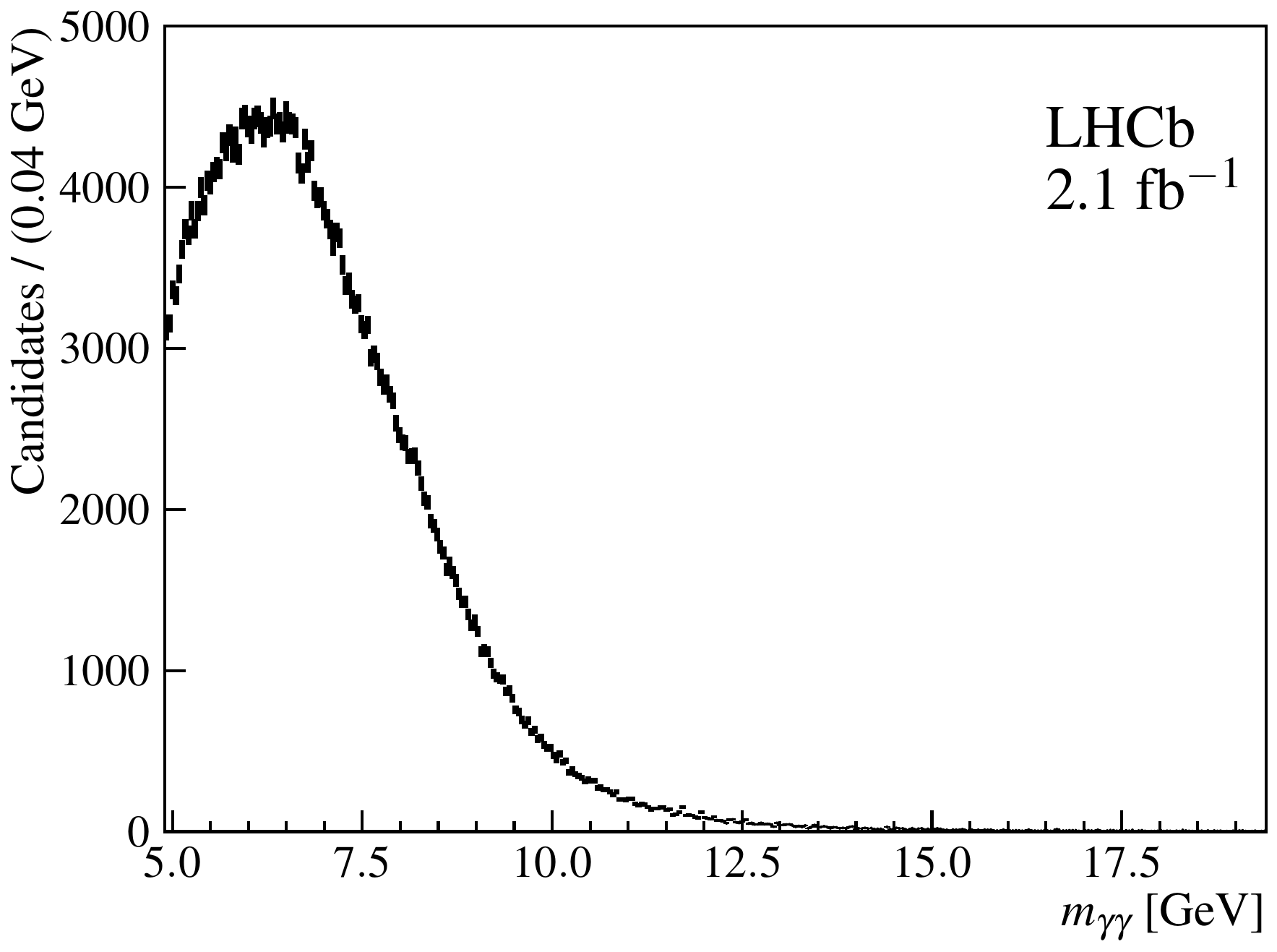}  \includegraphics[width=0.49\textwidth]{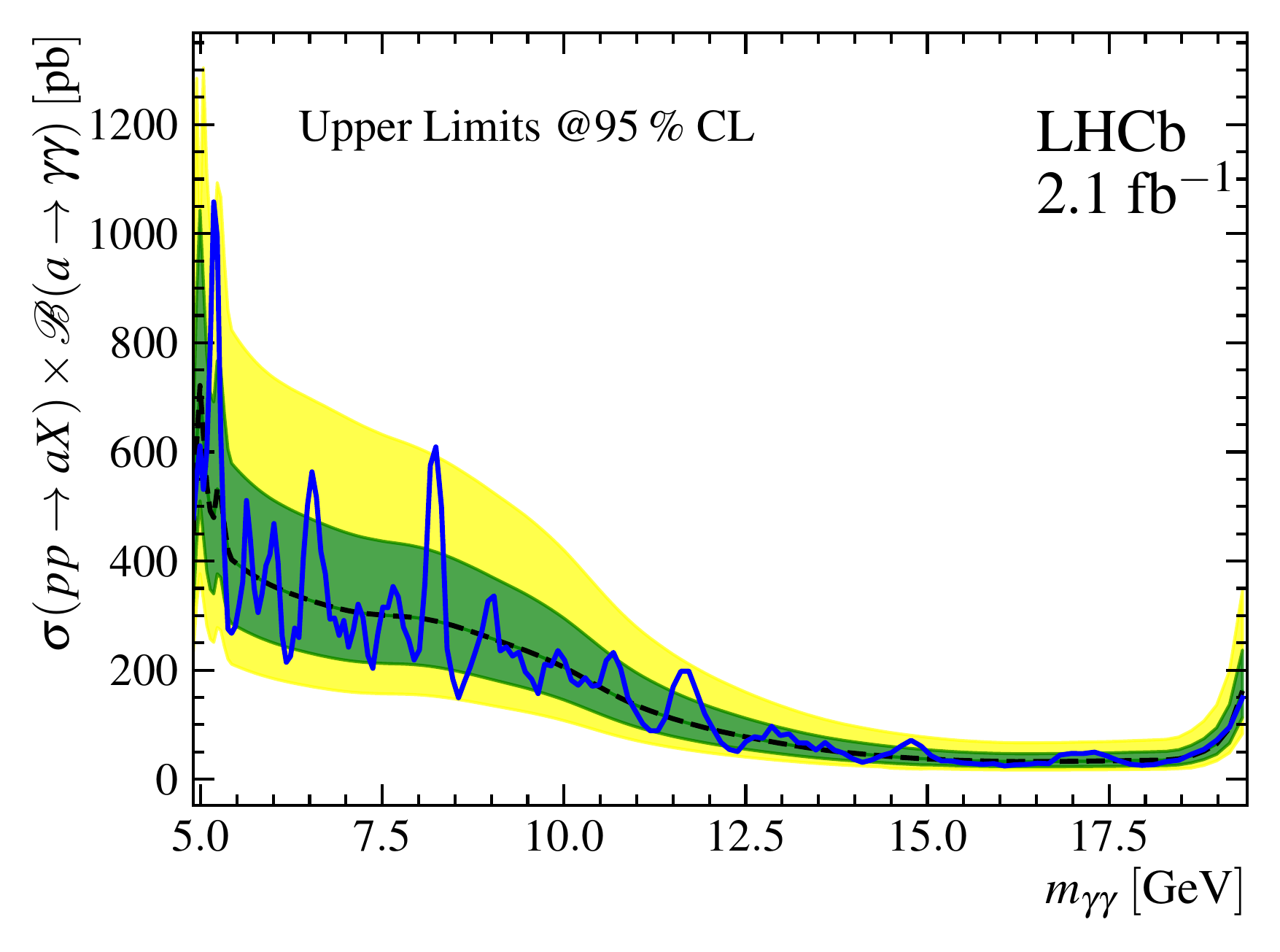}
  \caption{(Left) Spectrum of the fully selected photon-pair mass and (right) distribution of the upper limits on the value of cross-section times photon-pair decay branching fraction of ALPs. The observed upper limits are denoted with the blue line, the black dashed lines and bands represent the expected limits.} 
  \label{fig:ex_bumphunt_limit}
  \end{figure}

Values for the signal branching fractions multiplied by the production cross-sections are directly obtained from unbinned maximum-likelihood fits~\cite{py:zfit:2020} to the photon-pair mass. The mass spectrum, illustrated in Fig.~\ref{fig:ex_bumphunt_limit}, is scanned in steps of size \mbox{$\sigma(\aa)/2$}, where $\sigma(\aa$) is the photon-pair mass resolution of about $2.0\,\%$ relative to the inspected mass.
Each fit covers a range of $10 \sigma$ around the ALP mass hypothesis. 
The mass line shape of the signal decays is modelled with a double-sided Crystal Ball function~\cite{Skwarnicki:1986xj}, with shape parameters fixed from fits to simulation. The width of the Gaussian core is scaled with correction factors obtained from $\Bs \to \phi \gamma$ decays in data and simulation.
Uncertainties from the efficiency and luminosity are taken into account by leaving the values floating within a Gaussian constraint in the fit.

The remaining background after imposing all selection requirements is dominated by random combinations of photons or $\piz\to\gamma\gamma$ decays that are detected as a single ECAL cluster.
Peaking components such as \mbox{$\Bd \to \piz \piz$} or \mbox{$\OneS \to \piz \piz \gamma $} are found to be negligible.
Therefore, the background shape is modelled using a Legendre polynomial including all odd modes up to fifth order, and even modes up to second order.
To stabilise the background estimate at the low-mass boundary, the control region with \mbox{$m(\g\g)<4.8\gev$} is used to interpolate the background shape  through a simultaneous fit to the two datasets.
  \begin{figure}
    \centering
    \includegraphics[width=0.49\textwidth]{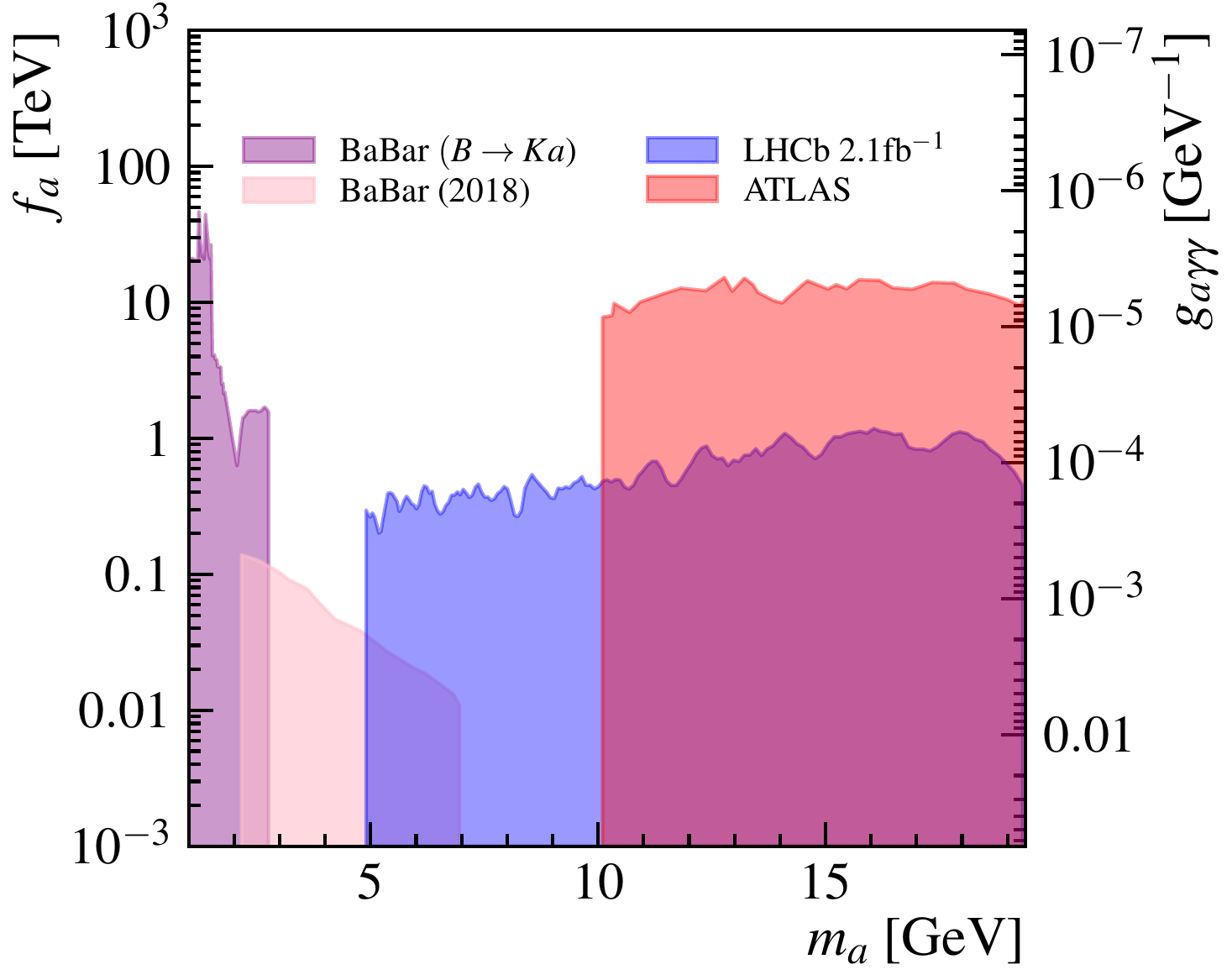}
    \caption{Projection of the limits shown in Fig.~\ref{fig:ex_bumphunt_limit} in the model discussed in Ref.~\cite{CidVidal:2018blh} and compared to data from Refs.~\cite{ATLAS:2022abz,BaBar:2011kau,btoka} assuming $c_1=c_2=c_3=10$. The shaded areas correspond to excluded regions at $95\,\%$ CL. On the right, the $y$-axis shows the ALP–photon coupling \mbox{$g_{a\gamma\gamma} \equiv \frac{\alpha_{\mathrm{EM}}(c_2 + 5/3 c_1)}{\pi f_a}$}, a standard QCD axion notation. The measurement reported in this Letter covers an area inaccessible to other experiments.}
    \label{fig:ex_bumphunt_limit_interpret}
    \end{figure}
No significant excess is found in the data and upper limits at $95\,\%$ confidence level (CL) are determined with the CL$_s$ method~\cite{CLs} using a one-sided test statistic~\cite{Cowan:2010js} as implemented in the hepstats software package~\cite{py:hepstats} in the mass range \mbox{[4.9,19.4]\gev}. The resulting limits are given in Fig.~\ref{fig:ex_bumphunt_limit}. The limit on the signal cross-section times branching fraction is interpreted as $1/f^2_a$ times a function of the ALP mass as described in Ref.~\cite{CidVidal:2018blh}. The upper limit on the cross-section results in a lower limit on $f_a$, which is shown in Fig.~\ref{fig:ex_bumphunt_limit_interpret} and compared to the results from Refs.~\cite{ATLAS:2022abz,BaBar:2011kau,btoka} for the typical choice of the $c_i$ coefficients $c_1=c_2=c_3=10$.

In parallel, separate dedicated fits are also performed to search for the decays of the  \Bd, \Bs and $\eta_b$ mesons into photon pairs at their known masses, following the same fitting strategy, where the efficiencies are taken from their respective corrected simulation.
For the measurement of the \mbox{$\Bds\to\gamma\gamma$} branching fractions, data selected with the dedicated classifier are used in the fit.
The observed (expected, for zero signal) upper limits are 
\begin{align*}
\mathcal{B}(\Bs\to\gamma\gamma)&<2.7\,(2.7) \times 10^{-5},\\
\mathcal{B}(\Bd\to\gamma\gamma) &<0.83\,(0.79) \times 10^{-5}\text{, and}\\\sigma(pp\to\eta_b X)\times\mathcal{B}(\eta_b\to\gamma\gamma) &<765 \,(710)\,\pb
\end{align*}

For ease of comparison with theoretical calculations, the result for \mbox{$\eta_b\to\gamma\gamma$} is also reported within a fiducial region of \mbox{$\pt(\eta_b)>6\gev$} and \mbox{$\eta(\eta_b)\in[1.8,5.2]$}, yielding \mbox{$\sigma(pp\to\eta_b X)\times\mathcal{B}(\eta_b\to\gamma\gamma)<84\pb$} at $95\,\%$ CL. The upper limits on the ALP production cross-section times branching fraction are also determined in the fiducial region and provided in the Supplemental Material to this Letter~\cite{supplemental}.

In summary, a search for resonances decaying to photon pairs in the mass range \mbox{$[4.9, 19.4]\gev$} is performed using proton-proton collision data recorded with the LHCb detector in 2018.
No excess is found, and upper limits on the branching fraction multiplied by the ALP production cross-section are set with sensitivities ranging from $\mathcal{O}(10)\,$pb to $\mathcal{O}(1000)\,\text{pb}$.
Applying a similar selection, upper limits on the branching fractions of \mbox{$\Bds\to\gamma\gamma$} decays and the product of the production cross-section and branching fraction of the decays of $\eta_b$ mesons to photon pairs are obtained.
This is the first search for a purely neutral final state with the LHCb experiment, bridging a large gap for ALP production in the mass between $4.9$ and $10.0\gev$, uniquely accessible to LHCb.

\clearpage
\section*{Supplemental material}

\begin{figure}[htbp]
    \centering
    \includegraphics[width=0.49\textwidth]{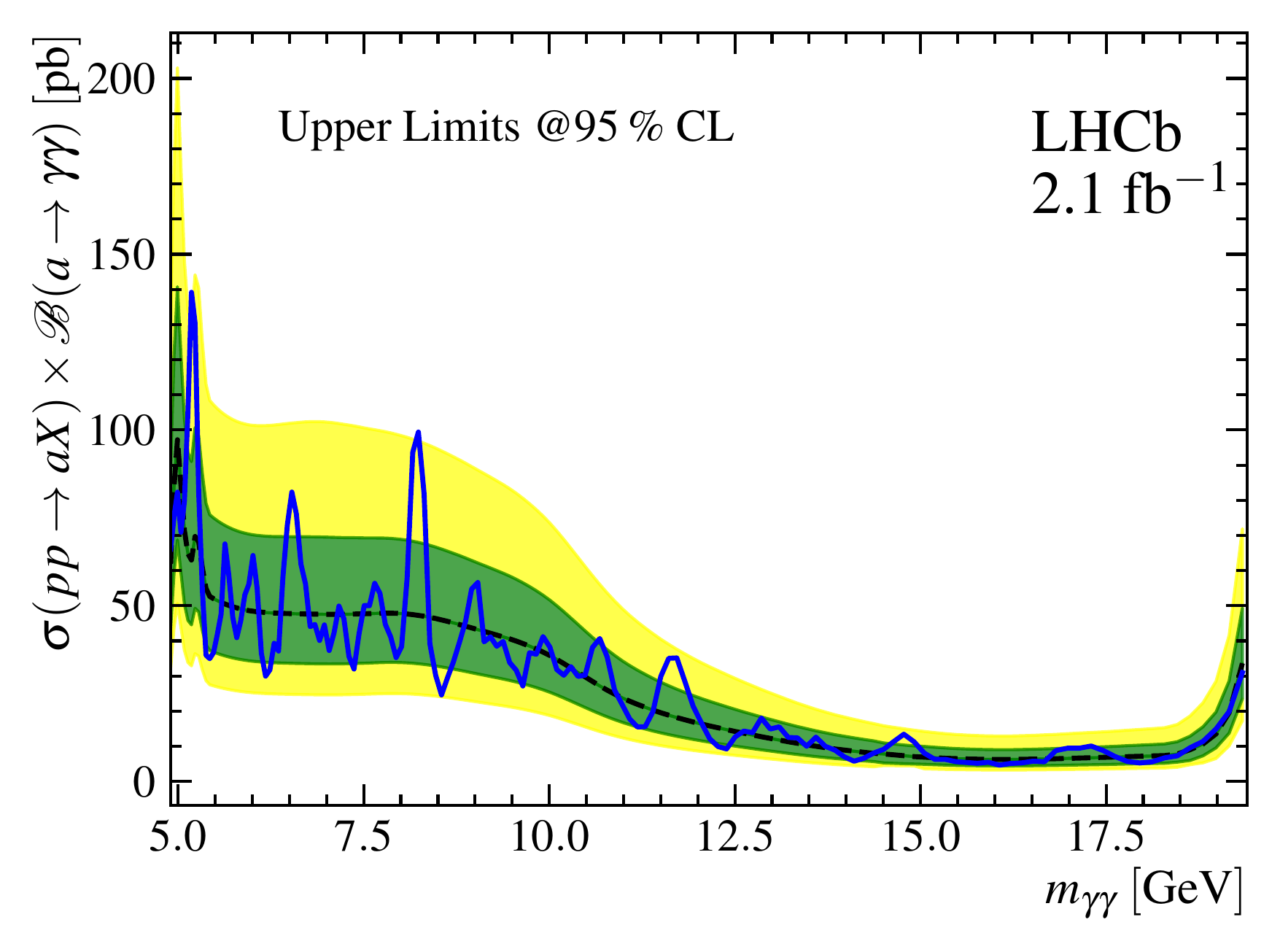}
    \caption{Upper limits on $\sigma(pp\to aX)\times\mathcal{B}(a\to\gamma\gamma)$ within the fiducial region of the analysis, $\pt(a)>6\,\gev$ and $\eta(a)\in[1.8,5.2]$.}
    \label{fig:ex_bumphunt_limit_fid}
    \end{figure}

In order to facilitate the reinterpretation in other models of physics beyond the SM, the upper limits on the cross-section times branching fraction of $a\to\gamma\gamma$ decays are additionally determined in a fiducial region, defined as $\pt(a)>6\,\gev$ and $\eta(a)\in[1.8,5.2]$. These upper limits are given in Fig.~\ref{fig:ex_bumphunt_limit_fid}. 

\section*{Acknowledgements}
\noindent We gratefully acknowledge Alberto Mariotti, Diego Redigolo, Filippo Sala, and Kohsaku Tobioka  for many useful discussions during the preparation of the analysis 
 and for their help with the theoretical interpretation of our measurement.
\noindent We express our gratitude to our colleagues in the CERN
accelerator departments for the excellent performance of the LHC. We
thank the technical and administrative staff at the LHCb
institutes.
We acknowledge support from CERN and from the national agencies:
ARC (Australia);
CAPES, CNPq, FAPERJ and FINEP (Brazil); 
MOST and NSFC (China); 
CNRS/IN2P3 (France); 
BMBF, DFG and MPG (Germany); 
INFN (Italy); 
NWO (Netherlands); 
MNiSW and NCN (Poland); 
MCID/IFA (Romania); 
MICIU and AEI (Spain);
SNSF and SER (Switzerland); 
NASU (Ukraine); 
STFC (United Kingdom); 
DOE NP and NSF (USA).
We acknowledge the computing resources that are provided by ARDC (Australia), 
CBPF (Brazil),
CERN, 
IHEP and LZU (China),
IN2P3 (France), 
KIT and DESY (Germany), 
INFN (Italy), 
SURF (Netherlands),
Polish WLCG (Poland),
IFIN-HH (Romania), 
PIC (Spain), CSCS (Switzerland), 
and GridPP (United Kingdom).
We are indebted to the communities behind the multiple open-source
software packages on which we depend.
Individual groups or members have received support from
Key Research Program of Frontier Sciences of CAS, CAS PIFI, CAS CCEPP, 
Fundamental Research Funds for the Central Universities,  and Sci.\ \& Tech.\ Program of Guangzhou (China);
Minciencias (Colombia);
EPLANET, Marie Sk\l{}odowska-Curie Actions, ERC and NextGenerationEU (European Union);
A*MIDEX, ANR, IPhU and Labex P2IO, and R\'{e}gion Auvergne-Rh\^{o}ne-Alpes (France);
Alexander-von-Humboldt Foundation (Germany);
ICSC (Italy); 
Severo Ochoa and Mar\'ia de Maeztu Units of Excellence, GVA, XuntaGal, GENCAT, InTalent-Inditex and Prog.~Atracci\'on Talento CM (Spain);
SRC (Sweden);
the Leverhulme Trust, the Royal Society and UKRI (United Kingdom).

\addcontentsline{toc}{section}{References}
\bibliographystyle{LHCb}
\bibliography{main,standard,LHCb-PAPER,LHCb-CONF,LHCb-DP,LHCb-TDR,alps_exp}

\ifx\mcitethebibliography\mciteundefinedmacro
\PackageError{LHCb.bst}{mciteplus.sty has not been loaded}
{This bibstyle requires the use of the mciteplus package.}\fi
\providecommand{\href}[2]{#2}
\begin{mcitethebibliography}{10}
\mciteSetBstSublistMode{n}
\mciteSetBstMaxWidthForm{subitem}{\alph{mcitesubitemcount})}
\mciteSetBstSublistLabelBeginEnd{\mcitemaxwidthsubitemform\space}
{\relax}{\relax}

\bibitem{Holdom:1982ex}
B.~Holdom and M.~E. Peskin, \ifthenelse{\boolean{articletitles}}{\emph{{Raising
  the axion mass}},
  }{}\href{https://doi.org/10.1016/0550-3213(82)90228-0}{Nucl.\ Phys.\
  \textbf{B208} (1982) 397}\relax
\mciteBstWouldAddEndPuncttrue
\mciteSetBstMidEndSepPunct{\mcitedefaultmidpunct}
{\mcitedefaultendpunct}{\mcitedefaultseppunct}\relax
\EndOfBibitem
\bibitem{Choi:1988sy}
K.~Choi, C.~W. Kim, and W.~K. Sze,
  \ifthenelse{\boolean{articletitles}}{\emph{{Mass renormalization by
  instantons and the strong {CP} problem}},
  }{}\href{https://doi.org/10.1103/PhysRevLett.61.794}{Phys.\ Rev.\ Lett.\
  \textbf{61} (1988) 794}\relax
\mciteBstWouldAddEndPuncttrue
\mciteSetBstMidEndSepPunct{\mcitedefaultmidpunct}
{\mcitedefaultendpunct}{\mcitedefaultseppunct}\relax
\EndOfBibitem
\bibitem{Holdom:1985vx}
B.~Holdom, \ifthenelse{\boolean{articletitles}}{\emph{{Strong QCD at high
  energies and a heavy axion}},
  }{}\href{https://doi.org/10.1016/0370-2693(85)90371-5}{Phys.\ Lett.\
  \textbf{B154} (1985) 316}, Erratum
  \href{https://doi.org/10.1016/0370-2693(85)91643-0}{ibid.\   \textbf{B156}
  (1985) 452}\relax
\mciteBstWouldAddEndPuncttrue
\mciteSetBstMidEndSepPunct{\mcitedefaultmidpunct}
{\mcitedefaultendpunct}{\mcitedefaultseppunct}\relax
\EndOfBibitem
\bibitem{Dine:1986bg}
M.~Dine and N.~Seiberg, \ifthenelse{\boolean{articletitles}}{\emph{{String
  theory and the strong {CP} problem}},
  }{}\href{https://doi.org/10.1016/0550-3213(86)90043-X}{Nucl.\ Phys.\
  \textbf{B273} (1986) 109}\relax
\mciteBstWouldAddEndPuncttrue
\mciteSetBstMidEndSepPunct{\mcitedefaultmidpunct}
{\mcitedefaultendpunct}{\mcitedefaultseppunct}\relax
\EndOfBibitem
\bibitem{Flynn:1987rs}
J.~M. Flynn and L.~Randall, \ifthenelse{\boolean{articletitles}}{\emph{{A
  computation of the small instanton contribution to the axion potential}},
  }{}\href{https://doi.org/10.1016/0550-3213(87)90089-7}{Nucl.\ Phys.\
  \textbf{B293} (1987) 731}\relax
\mciteBstWouldAddEndPuncttrue
\mciteSetBstMidEndSepPunct{\mcitedefaultmidpunct}
{\mcitedefaultendpunct}{\mcitedefaultseppunct}\relax
\EndOfBibitem
\bibitem{Rubakov:1997vp}
V.~A. Rubakov, \ifthenelse{\boolean{articletitles}}{\emph{{Grand unification
  and heavy axion}}, }{}\href{https://doi.org/10.1134/1.567390}{JETP Lett.\
  \textbf{65} (1997) 621},
  \href{http://arxiv.org/abs/hep-ph/9703409}{{\normalfont\ttfamily
  arXiv:hep-ph/9703409}}\relax
\mciteBstWouldAddEndPuncttrue
\mciteSetBstMidEndSepPunct{\mcitedefaultmidpunct}
{\mcitedefaultendpunct}{\mcitedefaultseppunct}\relax
\EndOfBibitem
\bibitem{Choi:1998ep}
K.~Choi and H.~D. Kim, \ifthenelse{\boolean{articletitles}}{\emph{{Small
  instanton contribution to the axion potential in supersymmetric models}},
  }{}\href{https://doi.org/10.1103/PhysRevD.59.072001}{Phys.\ Rev.\
  \textbf{D59} (1999) 072001},
  \href{http://arxiv.org/abs/hep-ph/9809286}{{\normalfont\ttfamily
  arXiv:hep-ph/9809286}}\relax
\mciteBstWouldAddEndPuncttrue
\mciteSetBstMidEndSepPunct{\mcitedefaultmidpunct}
{\mcitedefaultendpunct}{\mcitedefaultseppunct}\relax
\EndOfBibitem
\bibitem{Berezhiani:2000gh}
Z.~Berezhiani, L.~Gianfagna, and M.~Giannotti,
  \ifthenelse{\boolean{articletitles}}{\emph{{Strong CP problem and mirror
  world: the Weinberg--Wilczek axion revisited}},
  }{}\href{https://doi.org/10.1016/S0370-2693(00)01392-7}{Phys.\ Lett.\
  \textbf{B500} (2001) 286},
  \href{http://arxiv.org/abs/hep-ph/0009290}{{\normalfont\ttfamily
  arXiv:hep-ph/0009290}}\relax
\mciteBstWouldAddEndPuncttrue
\mciteSetBstMidEndSepPunct{\mcitedefaultmidpunct}
{\mcitedefaultendpunct}{\mcitedefaultseppunct}\relax
\EndOfBibitem
\bibitem{Hook:2014cda}
A.~Hook, \ifthenelse{\boolean{articletitles}}{\emph{{Anomalous solutions to the
  strong CP problem}},
  }{}\href{https://doi.org/10.1103/PhysRevLett.114.141801}{Phys.\ Rev.\ Lett.\
  \textbf{114} (2015) 141801},
  \href{http://arxiv.org/abs/1411.3325}{{\normalfont\ttfamily
  arXiv:1411.3325}}\relax
\mciteBstWouldAddEndPuncttrue
\mciteSetBstMidEndSepPunct{\mcitedefaultmidpunct}
{\mcitedefaultendpunct}{\mcitedefaultseppunct}\relax
\EndOfBibitem
\bibitem{Fukuda:2015ana}
H.~Fukuda, K.~Harigaya, M.~Ibe, and T.~T. Yanagida,
  \ifthenelse{\boolean{articletitles}}{\emph{{Model of visible QCD axion}},
  }{}\href{https://doi.org/10.1103/PhysRevD.92.015021}{Phys.\ Rev.\
  \textbf{D92} (2015) 015021},
  \href{http://arxiv.org/abs/1504.06084}{{\normalfont\ttfamily
  arXiv:1504.06084}}\relax
\mciteBstWouldAddEndPuncttrue
\mciteSetBstMidEndSepPunct{\mcitedefaultmidpunct}
{\mcitedefaultendpunct}{\mcitedefaultseppunct}\relax
\EndOfBibitem
\bibitem{Dimopoulos:2016lvn}
S.~Dimopoulos, A.~Hook, J.~Huang, and G.~Marques-Tavares,
  \ifthenelse{\boolean{articletitles}}{\emph{{A collider observable QCD
  axion}}, }{}\href{https://doi.org/10.1007/JHEP11(2016)052}{JHEP \textbf{11}
  (2016) 052}, \href{http://arxiv.org/abs/1606.03097}{{\normalfont\ttfamily
  arXiv:1606.03097}}\relax
\mciteBstWouldAddEndPuncttrue
\mciteSetBstMidEndSepPunct{\mcitedefaultmidpunct}
{\mcitedefaultendpunct}{\mcitedefaultseppunct}\relax
\EndOfBibitem
\bibitem{Agrawal:2017ksf}
P.~Agrawal and K.~Howe, \ifthenelse{\boolean{articletitles}}{\emph{{Factoring
  the strong CP problem}},
  }{}\href{https://doi.org/10.1007/JHEP12(2018)029}{JHEP \textbf{12} (2018)
  029}, \href{http://arxiv.org/abs/1710.04213}{{\normalfont\ttfamily
  arXiv:1710.04213}}\relax
\mciteBstWouldAddEndPuncttrue
\mciteSetBstMidEndSepPunct{\mcitedefaultmidpunct}
{\mcitedefaultendpunct}{\mcitedefaultseppunct}\relax
\EndOfBibitem
\bibitem{Gaillard:2018xgk}
M.~K. Gaillard {\em et~al.}, \ifthenelse{\boolean{articletitles}}{\emph{{Color
  unified dynamical axion}},
  }{}\href{https://doi.org/10.1140/epjc/s10052-018-6396-6}{Eur.\ Phys.\ J.\
  \textbf{C78} (2018) 972},
  \href{http://arxiv.org/abs/1805.06465}{{\normalfont\ttfamily
  arXiv:1805.06465}}\relax
\mciteBstWouldAddEndPuncttrue
\mciteSetBstMidEndSepPunct{\mcitedefaultmidpunct}
{\mcitedefaultendpunct}{\mcitedefaultseppunct}\relax
\EndOfBibitem
\bibitem{Bishara:2017pfq}
F.~Bishara, J.~Brod, B.~Grinstein, and J.~Zupan,
  \ifthenelse{\boolean{articletitles}}{\emph{{From quarks to nucleons in dark
  matter direct detection}},
  }{}\href{https://doi.org/10.1007/JHEP11(2017)059}{JHEP \textbf{11} (2017)
  059}, \href{http://arxiv.org/abs/1707.06998}{{\normalfont\ttfamily
  arXiv:1707.06998}}\relax
\mciteBstWouldAddEndPuncttrue
\mciteSetBstMidEndSepPunct{\mcitedefaultmidpunct}
{\mcitedefaultendpunct}{\mcitedefaultseppunct}\relax
\EndOfBibitem
\bibitem{Goh:2008xz}
H.-S. Goh and M.~Ibe, \ifthenelse{\boolean{articletitles}}{\emph{{R-axion
  detection at LHC}},
  }{}\href{https://doi.org/10.1088/1126-6708/2009/03/049}{JHEP \textbf{03}
  (2009) 049}, \href{http://arxiv.org/abs/0810.5773}{{\normalfont\ttfamily
  arXiv:0810.5773}}\relax
\mciteBstWouldAddEndPuncttrue
\mciteSetBstMidEndSepPunct{\mcitedefaultmidpunct}
{\mcitedefaultendpunct}{\mcitedefaultseppunct}\relax
\EndOfBibitem
\bibitem{Mardon:2009gw}
J.~Mardon, Y.~Nomura, and J.~Thaler,
  \ifthenelse{\boolean{articletitles}}{\emph{{Cosmic signals from the hidden
  sector}}, }{}\href{https://doi.org/10.1103/PhysRevD.80.035013}{Phys.\ Rev.\
  \textbf{D80} (2009) 035013},
  \href{http://arxiv.org/abs/0905.3749}{{\normalfont\ttfamily
  arXiv:0905.3749}}\relax
\mciteBstWouldAddEndPuncttrue
\mciteSetBstMidEndSepPunct{\mcitedefaultmidpunct}
{\mcitedefaultendpunct}{\mcitedefaultseppunct}\relax
\EndOfBibitem
\bibitem{Fan:2010is}
J.~Fan, J.~Thaler, and L.-T. Wang,
  \ifthenelse{\boolean{articletitles}}{\emph{{Dark matter from dynamical SUSY
  breaking}}, }{}\href{https://doi.org/10.1007/JHEP06(2010)045}{JHEP
  \textbf{06} (2010) 045},
  \href{http://arxiv.org/abs/1004.0008}{{\normalfont\ttfamily
  arXiv:1004.0008}}\relax
\mciteBstWouldAddEndPuncttrue
\mciteSetBstMidEndSepPunct{\mcitedefaultmidpunct}
{\mcitedefaultendpunct}{\mcitedefaultseppunct}\relax
\EndOfBibitem
\bibitem{Bellazzini:2017neg}
B.~Bellazzini {\em et~al.},
  \ifthenelse{\boolean{articletitles}}{\emph{{$R$-axion at colliders}},
  }{}\href{https://doi.org/10.1103/PhysRevLett.119.141804}{Phys.\ Rev.\ Lett.\
  \textbf{119} (2017) 141804},
  \href{http://arxiv.org/abs/1702.02152}{{\normalfont\ttfamily
  arXiv:1702.02152}}\relax
\mciteBstWouldAddEndPuncttrue
\mciteSetBstMidEndSepPunct{\mcitedefaultmidpunct}
{\mcitedefaultendpunct}{\mcitedefaultseppunct}\relax
\EndOfBibitem
\bibitem{Ferretti:2013kya}
G.~Ferretti and D.~Karateev,
  \ifthenelse{\boolean{articletitles}}{\emph{{Fermionic UV completions of
  composite Higgs models}},
  }{}\href{https://doi.org/10.1007/JHEP03(2014)077}{JHEP \textbf{03} (2014)
  077}, \href{http://arxiv.org/abs/1312.5330}{{\normalfont\ttfamily
  arXiv:1312.5330}}\relax
\mciteBstWouldAddEndPuncttrue
\mciteSetBstMidEndSepPunct{\mcitedefaultmidpunct}
{\mcitedefaultendpunct}{\mcitedefaultseppunct}\relax
\EndOfBibitem
\bibitem{Belyaev:2015hgo}
A.~Belyaev {\em et~al.}, \ifthenelse{\boolean{articletitles}}{\emph{{Singlets
  in composite Higgs models in light of the LHC 750 GeV diphoton excess}},
  }{}\href{https://doi.org/10.1103/PhysRevD.94.015004}{Phys.\ Rev.\
  \textbf{D94} (2016) 015004},
  \href{http://arxiv.org/abs/1512.07242}{{\normalfont\ttfamily
  arXiv:1512.07242}}\relax
\mciteBstWouldAddEndPuncttrue
\mciteSetBstMidEndSepPunct{\mcitedefaultmidpunct}
{\mcitedefaultendpunct}{\mcitedefaultseppunct}\relax
\EndOfBibitem
\bibitem{Ferretti:2016upr}
G.~Ferretti, \ifthenelse{\boolean{articletitles}}{\emph{{Gauge theories of
  partial compositeness: scenarios for Run-II of the LHC}},
  }{}\href{https://doi.org/10.1007/JHEP06(2016)107}{JHEP \textbf{06} (2016)
  107}, \href{http://arxiv.org/abs/1604.06467}{{\normalfont\ttfamily
  arXiv:1604.06467}}\relax
\mciteBstWouldAddEndPuncttrue
\mciteSetBstMidEndSepPunct{\mcitedefaultmidpunct}
{\mcitedefaultendpunct}{\mcitedefaultseppunct}\relax
\EndOfBibitem
\bibitem{Mariotti:2017vtv}
A.~Mariotti, D.~Redigolo, F.~Sala, and K.~Tobioka,
  \ifthenelse{\boolean{articletitles}}{\emph{{New LHC bound on low-mass
  diphoton resonances}},
  }{}\href{https://doi.org/10.1016/j.physletb.2018.06.039}{Phys.\ Lett.\
  \textbf{B783} (2018) 13},
  \href{http://arxiv.org/abs/1710.01743}{{\normalfont\ttfamily
  arXiv:1710.01743}}\relax
\mciteBstWouldAddEndPuncttrue
\mciteSetBstMidEndSepPunct{\mcitedefaultmidpunct}
{\mcitedefaultendpunct}{\mcitedefaultseppunct}\relax
\EndOfBibitem
\bibitem{CidVidal:2018blh}
X.~Cid~Vidal {\em et~al.}, \ifthenelse{\boolean{articletitles}}{\emph{{New
  axion searches at flavor factories}},
  }{}\href{https://doi.org/10.1007/JHEP01(2019)113}{JHEP \textbf{01} (2019)
  113}, Erratum \href{https://doi.org/10.1007/JHEP06(2020)141}{ibid.\
  \textbf{06} (2020) 141},
  \href{http://arxiv.org/abs/1810.09452}{{\normalfont\ttfamily
  arXiv:1810.09452}}\relax
\mciteBstWouldAddEndPuncttrue
\mciteSetBstMidEndSepPunct{\mcitedefaultmidpunct}
{\mcitedefaultendpunct}{\mcitedefaultseppunct}\relax
\EndOfBibitem
\bibitem{ATLAS:2020hii}
ATLAS collaboration, G.~Aad {\em et~al.},
  \ifthenelse{\boolean{articletitles}}{\emph{{Measurement of light-by-light
  scattering and search for axion-like particles with 2.2 nb$^{-1}$ of Pb+Pb
  data with the ATLAS detector}},
  }{}\href{https://doi.org/10.1007/JHEP11(2021)050}{JHEP \textbf{03} (2021)
  243}, Erratum \href{https://doi.org/10.1007/JHEP11(2021)050}{ibid.\
  \textbf{11} (2021) 050},
  \href{http://arxiv.org/abs/2008.05355}{{\normalfont\ttfamily
  arXiv:2008.05355}}\relax
\mciteBstWouldAddEndPuncttrue
\mciteSetBstMidEndSepPunct{\mcitedefaultmidpunct}
{\mcitedefaultendpunct}{\mcitedefaultseppunct}\relax
\EndOfBibitem
\bibitem{ATLAS:2022abz}
ATLAS collaboration, G.~Aad {\em et~al.},
  \ifthenelse{\boolean{articletitles}}{\emph{{Search for boosted diphoton
  resonances in the 10 to 70 GeV mass range using 138\,fb$^{-1}$ of 13 TeV pp
  collisions with the ATLAS detector}},
  }{}\href{https://doi.org/10.1007/JHEP07(2023)155}{JHEP \textbf{07} (2023)
  155}, \href{http://arxiv.org/abs/2211.04172}{{\normalfont\ttfamily
  arXiv:2211.04172}}\relax
\mciteBstWouldAddEndPuncttrue
\mciteSetBstMidEndSepPunct{\mcitedefaultmidpunct}
{\mcitedefaultendpunct}{\mcitedefaultseppunct}\relax
\EndOfBibitem
\bibitem{CMS:2018erd}
CMS collaboration, A.~M. Sirunyan {\em et~al.},
  \ifthenelse{\boolean{articletitles}}{\emph{{Evidence for light-by-light
  scattering and searches for axion-like particles in ultraperipheral PbPb
  collisions at $\sqrt{s_\mathrm{NN}} =$ 5.02 TeV}},
  }{}\href{https://doi.org/10.1016/j.physletb.2019.134826}{Phys.\ Lett.\
  \textbf{B797} (2019) 134826},
  \href{http://arxiv.org/abs/1810.04602}{{\normalfont\ttfamily
  arXiv:1810.04602}}\relax
\mciteBstWouldAddEndPuncttrue
\mciteSetBstMidEndSepPunct{\mcitedefaultmidpunct}
{\mcitedefaultendpunct}{\mcitedefaultseppunct}\relax
\EndOfBibitem
\bibitem{Dolan:2017osp}
M.~J. Dolan {\em et~al.}, \ifthenelse{\boolean{articletitles}}{\emph{{Revised
  constraints and Belle II sensitivity for visible and invisible axion-like
  particles}}, }{}\href{https://doi.org/10.1007/JHEP12(2017)094}{JHEP
  \textbf{12} (2017) 094}, Erratum
  \href{https://doi.org/10.1007/JHEP03(2021)190}{ibid.\   \textbf{03} (2021)
  190}, \href{http://arxiv.org/abs/1709.00009}{{\normalfont\ttfamily
  arXiv:1709.00009}}\relax
\mciteBstWouldAddEndPuncttrue
\mciteSetBstMidEndSepPunct{\mcitedefaultmidpunct}
{\mcitedefaultendpunct}{\mcitedefaultseppunct}\relax
\EndOfBibitem
\bibitem{CHARM:1985anb}
CHARM collaboration, F.~Bergsma {\em et~al.},
  \ifthenelse{\boolean{articletitles}}{\emph{{Search for axion like particle
  production in 400{GeV} proton-copper interactions}},
  }{}\href{https://doi.org/10.1016/0370-2693(85)90400-9}{Phys.\ Lett.\
  \textbf{B157} (1985) 458}\relax
\mciteBstWouldAddEndPuncttrue
\mciteSetBstMidEndSepPunct{\mcitedefaultmidpunct}
{\mcitedefaultendpunct}{\mcitedefaultseppunct}\relax
\EndOfBibitem
\bibitem{Riordan:1987aw}
E.~M. Riordan {\em et~al.}, \ifthenelse{\boolean{articletitles}}{\emph{{A
  search for short-lived axions in an electron-beam-dump experiment}},
  }{}\href{https://doi.org/10.1103/PhysRevLett.59.755}{Phys.\ Rev.\ Lett.\
  \textbf{59} (1987) 755}\relax
\mciteBstWouldAddEndPuncttrue
\mciteSetBstMidEndSepPunct{\mcitedefaultmidpunct}
{\mcitedefaultendpunct}{\mcitedefaultseppunct}\relax
\EndOfBibitem
\bibitem{Blumlein:1990ay}
J.~Blumlein {\em et~al.}, \ifthenelse{\boolean{articletitles}}{\emph{{Limits on
  neutral light scalar and pseudoscalar particles in a proton beam dump
  experiment}}, }{}\href{https://doi.org/10.1007/BF01548556}{Z.\ Phys.\
  \textbf{C51} (1991) 341}\relax
\mciteBstWouldAddEndPuncttrue
\mciteSetBstMidEndSepPunct{\mcitedefaultmidpunct}
{\mcitedefaultendpunct}{\mcitedefaultseppunct}\relax
\EndOfBibitem
\bibitem{NA64:2020qwq}
NA64 collaboration, D.~Banerjee {\em et~al.},
  \ifthenelse{\boolean{articletitles}}{\emph{{Search for axionlike and scalar
  particles with the NA64 experiment}},
  }{}\href{https://doi.org/10.1103/PhysRevLett.125.081801}{Phys.\ Rev.\ Lett.\
  \textbf{125} (2020) 081801},
  \href{http://arxiv.org/abs/2005.02710}{{\normalfont\ttfamily
  arXiv:2005.02710}}\relax
\mciteBstWouldAddEndPuncttrue
\mciteSetBstMidEndSepPunct{\mcitedefaultmidpunct}
{\mcitedefaultendpunct}{\mcitedefaultseppunct}\relax
\EndOfBibitem
\bibitem{Belle-II:2020jti}
Belle II collaboration, F.~Abudin\'en {\em et~al.},
  \ifthenelse{\boolean{articletitles}}{\emph{{Search for axion-like particles
  produced in $e^+e^-$ collisions at Belle II}},
  }{}\href{https://doi.org/10.1103/PhysRevLett.125.161806}{Phys.\ Rev.\ Lett.\
  \textbf{125} (2020) 161806},
  \href{http://arxiv.org/abs/2007.13071}{{\normalfont\ttfamily
  arXiv:2007.13071}}\relax
\mciteBstWouldAddEndPuncttrue
\mciteSetBstMidEndSepPunct{\mcitedefaultmidpunct}
{\mcitedefaultendpunct}{\mcitedefaultseppunct}\relax
\EndOfBibitem
\bibitem{BESIII:2022rzz}
BESIII collaboration, M.~Ablikim {\em et~al.},
  \ifthenelse{\boolean{articletitles}}{\emph{{Search for an axion-like particle
  in radiative J/\ensuremath{\psi} decays}},
  }{}\href{https://doi.org/10.1016/j.physletb.2023.137698}{Phys.\ Lett.\
  \textbf{B838} (2023) 137698},
  \href{http://arxiv.org/abs/2211.12699}{{\normalfont\ttfamily
  arXiv:2211.12699}}\relax
\mciteBstWouldAddEndPuncttrue
\mciteSetBstMidEndSepPunct{\mcitedefaultmidpunct}
{\mcitedefaultendpunct}{\mcitedefaultseppunct}\relax
\EndOfBibitem
\bibitem{BESIII:2024hdv}
BESIII collaboration, M.~Ablikim {\em et~al.},
  \ifthenelse{\boolean{articletitles}}{\emph{{Search for diphoton decays of an
  axionlike particle in radiative J/\ensuremath{\psi} decays}},
  }{}\href{https://doi.org/10.1103/PhysRevD.110.L031101}{Phys.\ Rev.\
  \textbf{D110} (2024) L031101},
  \href{http://arxiv.org/abs/2404.04640}{{\normalfont\ttfamily
  arXiv:2404.04640}}\relax
\mciteBstWouldAddEndPuncttrue
\mciteSetBstMidEndSepPunct{\mcitedefaultmidpunct}
{\mcitedefaultendpunct}{\mcitedefaultseppunct}\relax
\EndOfBibitem
\bibitem{Halliday:2024lca}
J.~W.~D. Halliday {\em et~al.},
  \ifthenelse{\boolean{articletitles}}{\emph{{Bounds on heavy axions with an
  X-ray free electron laser}},
  }{}\href{https://doi.org/10.1103/PhysRevLett.134.055001}{Phys.\ Rev.\ Lett.\
  \textbf{134} (2025) 055001},
  \href{http://arxiv.org/abs/2404.17333}{{\normalfont\ttfamily
  arXiv:2404.17333}}\relax
\mciteBstWouldAddEndPuncttrue
\mciteSetBstMidEndSepPunct{\mcitedefaultmidpunct}
{\mcitedefaultendpunct}{\mcitedefaultseppunct}\relax
\EndOfBibitem
\bibitem{FASER:2024bbl}
FASER collaboration, R.~Mammen~Abraham {\em et~al.},
  \ifthenelse{\boolean{articletitles}}{\emph{{Shining light on the dark sector:
  search for axion-like particles and other new physics in photonic final
  states with FASER}}, }{}\href{https://doi.org/10.1007/JHEP01(2025)199}{JHEP
  \textbf{01} (2025) 199},
  \href{http://arxiv.org/abs/2410.10363}{{\normalfont\ttfamily
  arXiv:2410.10363}}\relax
\mciteBstWouldAddEndPuncttrue
\mciteSetBstMidEndSepPunct{\mcitedefaultmidpunct}
{\mcitedefaultendpunct}{\mcitedefaultseppunct}\relax
\EndOfBibitem
\bibitem{Jaeckel:2015jla}
J.~Jaeckel and M.~Spannowsky,
  \ifthenelse{\boolean{articletitles}}{\emph{{Probing MeV to 90 GeV axion-like
  particles with LEP and LHC}},
  }{}\href{https://doi.org/10.1016/j.physletb.2015.12.037}{Phys.\ Lett.\
  \textbf{B753} (2016) 482},
  \href{http://arxiv.org/abs/1509.00476}{{\normalfont\ttfamily
  arXiv:1509.00476}}\relax
\mciteBstWouldAddEndPuncttrue
\mciteSetBstMidEndSepPunct{\mcitedefaultmidpunct}
{\mcitedefaultendpunct}{\mcitedefaultseppunct}\relax
\EndOfBibitem
\bibitem{Knapen:2016moh}
S.~Knapen, T.~Lin, H.~K. Lou, and T.~Melia,
  \ifthenelse{\boolean{articletitles}}{\emph{{Searching for axionlike particles
  with ultraperipheral heavy-ion collisions}},
  }{}\href{https://doi.org/10.1103/PhysRevLett.118.171801}{Phys.\ Rev.\ Lett.\
  \textbf{118} (2017) 171801},
  \href{http://arxiv.org/abs/1607.06083}{{\normalfont\ttfamily
  arXiv:1607.06083}}\relax
\mciteBstWouldAddEndPuncttrue
\mciteSetBstMidEndSepPunct{\mcitedefaultmidpunct}
{\mcitedefaultendpunct}{\mcitedefaultseppunct}\relax
\EndOfBibitem
\bibitem{Capozzi:2023ffu}
F.~Capozzi {\em et~al.}, \ifthenelse{\boolean{articletitles}}{\emph{{New
  constraints on ALP couplings to electrons and photons from ArgoNeuT and the
  MiniBooNE beam dump}},
  }{}\href{https://doi.org/10.1103/PhysRevD.108.075019}{Phys.\ Rev.\
  \textbf{D108} (2023) 075019},
  \href{http://arxiv.org/abs/2307.03878}{{\normalfont\ttfamily
  arXiv:2307.03878}}\relax
\mciteBstWouldAddEndPuncttrue
\mciteSetBstMidEndSepPunct{\mcitedefaultmidpunct}
{\mcitedefaultendpunct}{\mcitedefaultseppunct}\relax
\EndOfBibitem
\bibitem{NOMAD:2000usb}
NOMAD collaboration, P.~Astier {\em et~al.},
  \ifthenelse{\boolean{articletitles}}{\emph{{Search for eV (pseudo)scalar
  penetrating particles in the SPS neutrino beam}},
  }{}\href{https://doi.org/10.1016/S0370-2693(00)00375-0}{Phys.\ Lett.\
  \textbf{B479} (2000) 371}\relax
\mciteBstWouldAddEndPuncttrue
\mciteSetBstMidEndSepPunct{\mcitedefaultmidpunct}
{\mcitedefaultendpunct}{\mcitedefaultseppunct}\relax
\EndOfBibitem
\bibitem{PrimEx:2010fvg}
PrimEx collaboration, I.~Larin {\em et~al.},
  \ifthenelse{\boolean{articletitles}}{\emph{{A new measurement of the $\pi^0$
  radiative decay width}},
  }{}\href{https://doi.org/10.1103/PhysRevLett.106.162303}{Phys.\ Rev.\ Lett.\
  \textbf{106} (2011) 162303},
  \href{http://arxiv.org/abs/1009.1681}{{\normalfont\ttfamily
  arXiv:1009.1681}}\relax
\mciteBstWouldAddEndPuncttrue
\mciteSetBstMidEndSepPunct{\mcitedefaultmidpunct}
{\mcitedefaultendpunct}{\mcitedefaultseppunct}\relax
\EndOfBibitem
\bibitem{Aloni:2019ruo}
D.~Aloni, C.~Fanelli, Y.~Soreq, and M.~Williams,
  \ifthenelse{\boolean{articletitles}}{\emph{{Photoproduction of axionlike
  particles}}, }{}\href{https://doi.org/10.1103/PhysRevLett.123.071801}{Phys.\
  Rev.\ Lett.\  \textbf{123} (2019) 071801},
  \href{http://arxiv.org/abs/1903.03586}{{\normalfont\ttfamily
  arXiv:1903.03586}}\relax
\mciteBstWouldAddEndPuncttrue
\mciteSetBstMidEndSepPunct{\mcitedefaultmidpunct}
{\mcitedefaultendpunct}{\mcitedefaultseppunct}\relax
\EndOfBibitem
\bibitem{Pybus:2023yex}
J.~R. Pybus {\em et~al.}, \ifthenelse{\boolean{articletitles}}{\emph{{Search
  for axion-like particles through nuclear Primakoff production using the GlueX
  detector}}, }{}\href{https://doi.org/10.1016/j.physletb.2024.138790}{Phys.\
  Lett.\  \textbf{B855} (2024) 138790},
  \href{http://arxiv.org/abs/2308.06339}{{\normalfont\ttfamily
  arXiv:2308.06339}}\relax
\mciteBstWouldAddEndPuncttrue
\mciteSetBstMidEndSepPunct{\mcitedefaultmidpunct}
{\mcitedefaultendpunct}{\mcitedefaultseppunct}\relax
\EndOfBibitem
\bibitem{CMS-PAS-HIG-24-014}
CMS collaboration, \ifthenelse{\boolean{articletitles}}{\emph{{Search for a
  very low mass diphoton resonance using 2018 data}}, }{}
  \href{http://cdsweb.cern.ch/search?p=CMS-PAS-HIG-24-014&f=reportnumber&action_search=Search&c=LHCb}
  {CMS-PAS-HIG-24-014}, 2025\relax
\mciteBstWouldAddEndPuncttrue
\mciteSetBstMidEndSepPunct{\mcitedefaultmidpunct}
{\mcitedefaultendpunct}{\mcitedefaultseppunct}\relax
\EndOfBibitem
\bibitem{Belle-II:2025bkt}
Belle-II, Belle collaboration, I.~Adachi {\em et~al.},
  \ifthenelse{\boolean{articletitles}}{\emph{{Search for an axion-Like particle
  in $B\rightarrow K^{(*)} a (\rightarrow \gamma\gamma)$ decays at Belle}},
  }{}\href{http://arxiv.org/abs/2507.01249}{{\normalfont\ttfamily
  arXiv:2507.01249}}\relax
\mciteBstWouldAddEndPuncttrue
\mciteSetBstMidEndSepPunct{\mcitedefaultmidpunct}
{\mcitedefaultendpunct}{\mcitedefaultseppunct}\relax
\EndOfBibitem
\bibitem{Gavela:2019cmq}
M.~B. Gavela, J.~M. No, V.~Sanz, and J.~F. de~Troc\'oniz,
  \ifthenelse{\boolean{articletitles}}{\emph{{Nonresonant searches for
  axion-like particles at the LHC}},
  }{}\href{https://doi.org/10.1103/PhysRevLett.124.051802}{Phys.\ Rev.\ Lett.\
  \textbf{124} (2020) 051802},
  \href{http://arxiv.org/abs/1905.12953}{{\normalfont\ttfamily
  arXiv:1905.12953}}\relax
\mciteBstWouldAddEndPuncttrue
\mciteSetBstMidEndSepPunct{\mcitedefaultmidpunct}
{\mcitedefaultendpunct}{\mcitedefaultseppunct}\relax
\EndOfBibitem
\bibitem{Aliev:1998va}
T.~M. Aliev and E.~O. Iltan,
  \ifthenelse{\boolean{articletitles}}{\emph{{$\Bds\to\gamma\gamma$ decay in
  the two Higgs doublet model with flavor changing neutral currents}},
  }{}\href{https://doi.org/10.1103/PhysRevD.58.095014}{Phys.\ Rev.\
  \textbf{D58} (1998) 095014},
  \href{http://arxiv.org/abs/hep-ph/9803459}{{\normalfont\ttfamily
  arXiv:hep-ph/9803459}}\relax
\mciteBstWouldAddEndPuncttrue
\mciteSetBstMidEndSepPunct{\mcitedefaultmidpunct}
{\mcitedefaultendpunct}{\mcitedefaultseppunct}\relax
\EndOfBibitem
\bibitem{Gemintern:2004bw}
A.~Gemintern, S.~Bar-Shalom, and G.~Eilam,
  \ifthenelse{\boolean{articletitles}}{\emph{{$\B\to X_{(s)} \gamma \gamma$ and
  $\Bds\to \gamma \gamma$ in supersymmetry with broken R parity}},
  }{}\href{https://doi.org/10.1103/PhysRevD.70.035008}{Phys.\ Rev.\
  \textbf{D70} (2004) 035008},
  \href{http://arxiv.org/abs/hep-ph/0404152}{{\normalfont\ttfamily
  arXiv:hep-ph/0404152}}\relax
\mciteBstWouldAddEndPuncttrue
\mciteSetBstMidEndSepPunct{\mcitedefaultmidpunct}
{\mcitedefaultendpunct}{\mcitedefaultseppunct}\relax
\EndOfBibitem
\bibitem{BaBar:2010qny}
BaBar collaboration, P.~del Amo~Sanchez {\em et~al.},
  \ifthenelse{\boolean{articletitles}}{\emph{{Search for the decay $B^{0} \to
  \gamma \gamma$}}, }{}\href{https://doi.org/10.1103/PhysRevD.83.032006}{Phys.\
  Rev.\  \textbf{D83} (2011) 032006},
  \href{http://arxiv.org/abs/1010.2229}{{\normalfont\ttfamily
  arXiv:1010.2229}}\relax
\mciteBstWouldAddEndPuncttrue
\mciteSetBstMidEndSepPunct{\mcitedefaultmidpunct}
{\mcitedefaultendpunct}{\mcitedefaultseppunct}\relax
\EndOfBibitem
\bibitem{Belle:2014sac}
Belle collaboration, D.~Dutta {\em et~al.},
  \ifthenelse{\boolean{articletitles}}{\emph{{Search for
  $B_{s}^{0}\rightarrow\gamma\gamma$ and a measurement of the branching
  fraction for $B_{s}^{0}\rightarrow\phi\gamma$}},
  }{}\href{https://doi.org/10.1103/PhysRevD.91.011101}{Phys.\ Rev.\
  \textbf{D91} (2015) 011101},
  \href{http://arxiv.org/abs/1411.7771}{{\normalfont\ttfamily
  arXiv:1411.7771}}\relax
\mciteBstWouldAddEndPuncttrue
\mciteSetBstMidEndSepPunct{\mcitedefaultmidpunct}
{\mcitedefaultendpunct}{\mcitedefaultseppunct}\relax
\EndOfBibitem
\bibitem{Belle:2024bsw}
Belle, Belle II collaboration, I.~Adachi {\em et~al.},
  \ifthenelse{\boolean{articletitles}}{\emph{{Search for the decay
  \Bd\textrightarrow{}\ensuremath{\gamma}\ensuremath{\gamma} using Belle and
  Belle II data}},
  }{}\href{https://doi.org/10.1103/PhysRevD.110.L031106}{Phys.\ Rev.\
  \textbf{D110} (2024) L031106},
  \href{http://arxiv.org/abs/2405.19734}{{\normalfont\ttfamily
  arXiv:2405.19734}}\relax
\mciteBstWouldAddEndPuncttrue
\mciteSetBstMidEndSepPunct{\mcitedefaultmidpunct}
{\mcitedefaultendpunct}{\mcitedefaultseppunct}\relax
\EndOfBibitem
\bibitem{Bosch:2002bv}
S.~W. Bosch and G.~Buchalla, \ifthenelse{\boolean{articletitles}}{\emph{{The
  double radiative decays $B \to\gamma \gamma$ in the heavy quark limit}},
  }{}\href{https://doi.org/10.1088/1126-6708/2002/08/054}{JHEP \textbf{08}
  (2002) 054}, \href{http://arxiv.org/abs/hep-ph/0208202}{{\normalfont\ttfamily
  arXiv:hep-ph/0208202}}\relax
\mciteBstWouldAddEndPuncttrue
\mciteSetBstMidEndSepPunct{\mcitedefaultmidpunct}
{\mcitedefaultendpunct}{\mcitedefaultseppunct}\relax
\EndOfBibitem
\bibitem{Shen:2020hfq}
Y.-L. Shen, Y.-M. Wang, and Y.-B. Wei,
  \ifthenelse{\boolean{articletitles}}{\emph{{Precision calculations of the
  double radiative bottom-meson decays in soft-collinear effective theory}},
  }{}\href{https://doi.org/10.1007/JHEP12(2020)169}{JHEP \textbf{12} (2020)
  169}, \href{http://arxiv.org/abs/2009.02723}{{\normalfont\ttfamily
  arXiv:2009.02723}}\relax
\mciteBstWouldAddEndPuncttrue
\mciteSetBstMidEndSepPunct{\mcitedefaultmidpunct}
{\mcitedefaultendpunct}{\mcitedefaultseppunct}\relax
\EndOfBibitem
\bibitem{Li:2022mwo}
D.-H. Li, L.-Y. Li, C.-D. L\"u, and Y.-L. Shen,
  \ifthenelse{\boolean{articletitles}}{\emph{{Strange quark mass effect in
  $\Bs\to\gamma\gamma, \gamma\ell\ell$ decays}},
  }{}\href{https://doi.org/10.1103/PhysRevD.106.094038}{Phys.\ Rev.\
  \textbf{D106} (2022) 094038},
  \href{http://arxiv.org/abs/2205.05528}{{\normalfont\ttfamily
  arXiv:2205.05528}}\relax
\mciteBstWouldAddEndPuncttrue
\mciteSetBstMidEndSepPunct{\mcitedefaultmidpunct}
{\mcitedefaultendpunct}{\mcitedefaultseppunct}\relax
\EndOfBibitem
\bibitem{PDG2024}
Particle Data Group, S.~Navas {\em et~al.},
  \ifthenelse{\boolean{articletitles}}{\emph{{\href{http://pdg.lbl.gov/}{Review
  of particle physics}}},
  }{}\href{https://doi.org/10.1103/PhysRevD.110.030001}{Phys.\ Rev.\
  \textbf{D110} (2024) 030001}\relax
\mciteBstWouldAddEndPuncttrue
\mciteSetBstMidEndSepPunct{\mcitedefaultmidpunct}
{\mcitedefaultendpunct}{\mcitedefaultseppunct}\relax
\EndOfBibitem
\bibitem{Lansberg:2020ejc}
J.-P. Lansberg and M.~A. Ozcelik,
  \ifthenelse{\boolean{articletitles}}{\emph{{Curing the unphysical behaviour
  of NLO quarkonium production at the LHC and its relevance to constrain the
  gluon PDF at low scales}},
  }{}\href{https://doi.org/10.1140/epjc/s10052-021-09258-7}{Eur.\ Phys.\ J.\
  \textbf{C81} (2021) 497},
  \href{http://arxiv.org/abs/2012.00702}{{\normalfont\ttfamily
  arXiv:2012.00702}}\relax
\mciteBstWouldAddEndPuncttrue
\mciteSetBstMidEndSepPunct{\mcitedefaultmidpunct}
{\mcitedefaultendpunct}{\mcitedefaultseppunct}\relax
\EndOfBibitem
\bibitem{Colquhoun:2024wsj}
B.~Colquhoun, C.~T.~H. Davies, and G.~P. Lepage,
  \ifthenelse{\boolean{articletitles}}{\emph{{Precise prediction of the decay
  rate for $\eta_b\to \gamma \gamma$ from lattice QCD}},
  }{}\href{http://arxiv.org/abs/2410.24041}{{\normalfont\ttfamily
  arXiv:2410.24041}}\relax
\mciteBstWouldAddEndPuncttrue
\mciteSetBstMidEndSepPunct{\mcitedefaultmidpunct}
{\mcitedefaultendpunct}{\mcitedefaultseppunct}\relax
\EndOfBibitem
\bibitem{ALEPH:2002xdz}
ALEPH collaboration, A.~Heister {\em et~al.},
  \ifthenelse{\boolean{articletitles}}{\emph{{Search for
  $\gamma\gamma\to\eta_{(b)}$ in \epem collisions at LEP-2}},
  }{}\href{https://doi.org/10.1016/S0370-2693(02)01329-1}{Phys.\ Lett.\
  \textbf{B530} (2002) 56},
  \href{http://arxiv.org/abs/hep-ex/0202011}{{\normalfont\ttfamily
  arXiv:hep-ex/0202011}}\relax
\mciteBstWouldAddEndPuncttrue
\mciteSetBstMidEndSepPunct{\mcitedefaultmidpunct}
{\mcitedefaultendpunct}{\mcitedefaultseppunct}\relax
\EndOfBibitem
\bibitem{Abdallah:2006yg}
DELPHI collaboration, J.~Abdallah,
  \ifthenelse{\boolean{articletitles}}{\emph{{Search for $\eta_{(b)}$ in
  two-photon collisions at LEP II with the DELPHI detector}},
  }{}\href{https://doi.org/10.1016/j.physletb.2006.01.058}{Phys.\ Lett.\
  \textbf{B634} (2006) 340},
  \href{http://arxiv.org/abs/hep-ex/0601042}{{\normalfont\ttfamily
  arXiv:hep-ex/0601042}}\relax
\mciteBstWouldAddEndPuncttrue
\mciteSetBstMidEndSepPunct{\mcitedefaultmidpunct}
{\mcitedefaultendpunct}{\mcitedefaultseppunct}\relax
\EndOfBibitem
\bibitem{LHCb-DP-2008-001}
LHCb collaboration, A.~A. Alves~Jr.\ {\em et~al.},
  \ifthenelse{\boolean{articletitles}}{\emph{{The \lhcb detector at the LHC}},
  }{}\href{https://doi.org/10.1088/1748-0221/3/08/S08005}{JINST \textbf{3}
  (2008) S08005}\relax
\mciteBstWouldAddEndPuncttrue
\mciteSetBstMidEndSepPunct{\mcitedefaultmidpunct}
{\mcitedefaultendpunct}{\mcitedefaultseppunct}\relax
\EndOfBibitem
\bibitem{LHCb-DP-2014-002}
LHCb collaboration, R.~Aaij {\em et~al.},
  \ifthenelse{\boolean{articletitles}}{\emph{{LHCb detector performance}},
  }{}\href{https://doi.org/10.1142/S0217751X15300227}{Int.\ J.\ Mod.\ Phys.\
  \textbf{A30} (2015) 1530022},
  \href{http://arxiv.org/abs/1412.6352}{{\normalfont\ttfamily
  arXiv:1412.6352}}\relax
\mciteBstWouldAddEndPuncttrue
\mciteSetBstMidEndSepPunct{\mcitedefaultmidpunct}
{\mcitedefaultendpunct}{\mcitedefaultseppunct}\relax
\EndOfBibitem
\bibitem{LHCb-DP-2020-001}
C.~Abellan~Beteta {\em et~al.},
  \ifthenelse{\boolean{articletitles}}{\emph{{Calibration and performance of
  the LHCb calorimeters in Run 1 and 2 at the LHC}},
  }{}\href{http://arxiv.org/abs/2008.11556}{{\normalfont\ttfamily
  arXiv:2008.11556}}, {submitted to JINST}\relax
\mciteBstWouldAddEndPuncttrue
\mciteSetBstMidEndSepPunct{\mcitedefaultmidpunct}
{\mcitedefaultendpunct}{\mcitedefaultseppunct}\relax
\EndOfBibitem
\bibitem{Sjostrand:2007gs}
T.~Sj\"{o}strand, S.~Mrenna, and P.~Skands,
  \ifthenelse{\boolean{articletitles}}{\emph{{A brief introduction to PYTHIA
  8.1}}, }{}\href{https://doi.org/10.1016/j.cpc.2008.01.036}{Comput.\ Phys.\
  Commun.\  \textbf{178} (2008) 852},
  \href{http://arxiv.org/abs/0710.3820}{{\normalfont\ttfamily
  arXiv:0710.3820}}\relax
\mciteBstWouldAddEndPuncttrue
\mciteSetBstMidEndSepPunct{\mcitedefaultmidpunct}
{\mcitedefaultendpunct}{\mcitedefaultseppunct}\relax
\EndOfBibitem
\bibitem{Lange:2001uf}
D.~J. Lange, \ifthenelse{\boolean{articletitles}}{\emph{{The EvtGen particle
  decay simulation package}},
  }{}\href{https://doi.org/10.1016/S0168-9002(01)00089-4}{Nucl.\ Instrum.\
  Meth.\  \textbf{A462} (2001) 152}\relax
\mciteBstWouldAddEndPuncttrue
\mciteSetBstMidEndSepPunct{\mcitedefaultmidpunct}
{\mcitedefaultendpunct}{\mcitedefaultseppunct}\relax
\EndOfBibitem
\bibitem{Agostinelli:2002hh}
Geant4 collaboration, S.~Agostinelli {\em et~al.},
  \ifthenelse{\boolean{articletitles}}{\emph{{Geant4: A simulation toolkit}},
  }{}\href{https://doi.org/10.1016/S0168-9002(03)01368-8}{Nucl.\ Instrum.\
  Meth.\  \textbf{A506} (2003) 250}\relax
\mciteBstWouldAddEndPuncttrue
\mciteSetBstMidEndSepPunct{\mcitedefaultmidpunct}
{\mcitedefaultendpunct}{\mcitedefaultseppunct}\relax
\EndOfBibitem
\bibitem{Allison:2006ve}
Geant4 collaboration, J.~Allison {\em et~al.},
  \ifthenelse{\boolean{articletitles}}{\emph{{Geant4 developments and
  applications}}, }{}\href{https://doi.org/10.1109/TNS.2006.869826}{IEEE
  Trans.\ Nucl.\ Sci.\  \textbf{53} (2006) 270}\relax
\mciteBstWouldAddEndPuncttrue
\mciteSetBstMidEndSepPunct{\mcitedefaultmidpunct}
{\mcitedefaultendpunct}{\mcitedefaultseppunct}\relax
\EndOfBibitem
\bibitem{LHCb-PROC-2011-006}
M.~Clemencic {\em et~al.}, \ifthenelse{\boolean{articletitles}}{\emph{{The
  \lhcb simulation application, Gauss: Design, evolution and experience}},
  }{}\href{https://doi.org/10.1088/1742-6596/331/3/032023}{J.\ Phys.\ Conf.\
  Ser.\  \textbf{331} (2011) 032023}\relax
\mciteBstWouldAddEndPuncttrue
\mciteSetBstMidEndSepPunct{\mcitedefaultmidpunct}
{\mcitedefaultendpunct}{\mcitedefaultseppunct}\relax
\EndOfBibitem
\bibitem{madgraph}
J.~Alwall {\em et~al.}, \ifthenelse{\boolean{articletitles}}{\emph{{The
  automated computation of tree-level and next-to-leading order differential
  cross sections, and their matching to parton shower simulations}},
  }{}\href{https://doi.org/10.1007/JHEP07(2014)079}{JHEP \textbf{07} (2014)
  079}, \href{http://arxiv.org/abs/1405.0301}{{\normalfont\ttfamily
  arXiv:1405.0301}}\relax
\mciteBstWouldAddEndPuncttrue
\mciteSetBstMidEndSepPunct{\mcitedefaultmidpunct}
{\mcitedefaultendpunct}{\mcitedefaultseppunct}\relax
\EndOfBibitem
\bibitem{NNPDF:2014otw}
NNPDF collaboration, R.~D. Ball {\em et~al.},
  \ifthenelse{\boolean{articletitles}}{\emph{{Parton distributions for the LHC
  Run II}}, }{}\href{https://doi.org/10.1007/JHEP04(2015)040}{JHEP \textbf{04}
  (2015) 040}, \href{http://arxiv.org/abs/1410.8849}{{\normalfont\ttfamily
  arXiv:1410.8849}}\relax
\mciteBstWouldAddEndPuncttrue
\mciteSetBstMidEndSepPunct{\mcitedefaultmidpunct}
{\mcitedefaultendpunct}{\mcitedefaultseppunct}\relax
\EndOfBibitem
\bibitem{rta-pairs}
S.~Benson, A.~Casais~Vidal, X.~Cid~Vidal, and A.~Puig~Navarro,
  \ifthenelse{\boolean{articletitles}}{\emph{{Real-time discrimination of
  photon pairs using machine learning at the LHC}},
  }{}\href{https://doi.org/10.21468/SciPostPhys.7.5.062}{SciPost Phys.\
  \textbf{7} (2019) 062},
  \href{http://arxiv.org/abs/1906.09058}{{\normalfont\ttfamily
  arXiv:1906.09058}}\relax
\mciteBstWouldAddEndPuncttrue
\mciteSetBstMidEndSepPunct{\mcitedefaultmidpunct}
{\mcitedefaultendpunct}{\mcitedefaultseppunct}\relax
\EndOfBibitem
\bibitem{xgboost}
T.~Chen and C.~Guestrin, \ifthenelse{\boolean{articletitles}}{\emph{{XGBoost}:
  a scalable tree boosting system}, }{} in {\em Proceedings of the 22nd ACM
  SIGKDD International Conference on Knowledge Discovery and Data Mining},
  \href{https://doi.org/10.1145/2939672.2939785}{ KDD '16, (New York, NY, USA),
  785--794, ACM, 2016}\relax
\mciteBstWouldAddEndPuncttrue
\mciteSetBstMidEndSepPunct{\mcitedefaultmidpunct}
{\mcitedefaultendpunct}{\mcitedefaultseppunct}\relax
\EndOfBibitem
\bibitem{LHCb-PAPER-2020-046}
LHCb collaboration, R.~Aaij {\em et~al.},
  \ifthenelse{\boolean{articletitles}}{\emph{{Precise measurement of the
  $f_s/f_d$ ratio of fragmentation fractions and of $B^0_s$ decay branching
  fractions}}, }{}\href{https://doi.org/10.1103/PhysRevD.104.032005}{Phys.\
  Rev.\  \textbf{D104} (2021) 032005},
  \href{http://arxiv.org/abs/2103.06810}{{\normalfont\ttfamily
  arXiv:2103.06810}}\relax
\mciteBstWouldAddEndPuncttrue
\mciteSetBstMidEndSepPunct{\mcitedefaultmidpunct}
{\mcitedefaultendpunct}{\mcitedefaultseppunct}\relax
\EndOfBibitem
\bibitem{py:zfit:2020}
J.~Eschle, A.~{Puig Navarro}, R.~{Silva Coutinho}, and N.~Serra,
  \ifthenelse{\boolean{articletitles}}{\emph{zfit: Scalable pythonic fitting},
  }{}\href{https://doi.org/https://doi.org/10.1016/j.softx.2020.100508}{SoftwareX
  \textbf{11} (2020) 100508}\relax
\mciteBstWouldAddEndPuncttrue
\mciteSetBstMidEndSepPunct{\mcitedefaultmidpunct}
{\mcitedefaultendpunct}{\mcitedefaultseppunct}\relax
\EndOfBibitem
\bibitem{Skwarnicki:1986xj}
T.~Skwarnicki, {\em {A study of the radiative cascade transitions between the
  Upsilon-prime and Upsilon resonances}}, PhD thesis, Institute of Nuclear
  Physics, Krakow, 1986,
  {\href{http://inspirehep.net/record/230779/}{DESY-F31-86-02}}\relax
\mciteBstWouldAddEndPuncttrue
\mciteSetBstMidEndSepPunct{\mcitedefaultmidpunct}
{\mcitedefaultendpunct}{\mcitedefaultseppunct}\relax
\EndOfBibitem
\bibitem{BaBar:2011kau}
BaBar collaboration, J.~P. Lees {\em et~al.},
  \ifthenelse{\boolean{articletitles}}{\emph{{Search for hadronic decays of a
  light Higgs boson in the radiative decay $\Upsilon \to \gamma A^0$}},
  }{}\href{https://doi.org/10.1103/PhysRevLett.107.221803}{Phys.\ Rev.\ Lett.\
  \textbf{107} (2011) 221803},
  \href{http://arxiv.org/abs/1108.3549}{{\normalfont\ttfamily
  arXiv:1108.3549}}\relax
\mciteBstWouldAddEndPuncttrue
\mciteSetBstMidEndSepPunct{\mcitedefaultmidpunct}
{\mcitedefaultendpunct}{\mcitedefaultseppunct}\relax
\EndOfBibitem
\bibitem{btoka}
S.~Chakraborty {\em et~al.}, \ifthenelse{\boolean{articletitles}}{\emph{{Heavy
  QCD axion in $b\to s$ transition: enhanced limits and projections}},
  }{}\href{https://doi.org/10.1103/PhysRevD.104.055036}{Phys.\ Rev.\
  \textbf{D104} (2021) 055036},
  \href{http://arxiv.org/abs/2102.04474}{{\normalfont\ttfamily
  arXiv:2102.04474}}\relax
\mciteBstWouldAddEndPuncttrue
\mciteSetBstMidEndSepPunct{\mcitedefaultmidpunct}
{\mcitedefaultendpunct}{\mcitedefaultseppunct}\relax
\EndOfBibitem
\bibitem{CLs}
A.~L. Read, \ifthenelse{\boolean{articletitles}}{\emph{{Presentation of search
  results: The CL$_{\rm s}$ technique}},
  }{}\href{https://doi.org/10.1088/0954-3899/28/10/313}{J.\ Phys.\
  \textbf{G28} (2002) 2693}\relax
\mciteBstWouldAddEndPuncttrue
\mciteSetBstMidEndSepPunct{\mcitedefaultmidpunct}
{\mcitedefaultendpunct}{\mcitedefaultseppunct}\relax
\EndOfBibitem
\bibitem{Cowan:2010js}
G.~Cowan, K.~Cranmer, E.~Gross, and O.~Vitells,
  \ifthenelse{\boolean{articletitles}}{\emph{{Asymptotic formulae for
  likelihood-based tests of new physics}},
  }{}\href{https://doi.org/10.1140/epjc/s10052-011-1554-0}{Eur.\ Phys.\ J.\
  \textbf{C71} (2011) 1554}, Erratum
  \href{https://doi.org/10.1140/epjc/s10052-013-2501-z}{ibid.\   \textbf{73}
  (2013) 2501}, \href{http://arxiv.org/abs/1007.1727}{{\normalfont\ttfamily
  arXiv:1007.1727}}\relax
\mciteBstWouldAddEndPuncttrue
\mciteSetBstMidEndSepPunct{\mcitedefaultmidpunct}
{\mcitedefaultendpunct}{\mcitedefaultseppunct}\relax
\EndOfBibitem
\bibitem{py:hepstats}
M.~Marinangeli {\em et~al.},
  \ifthenelse{\boolean{articletitles}}{\emph{{hepstats}}, }{}
\newblock
  doi:~\href{https://doi.org/10.5281/zenodo.3519200}{10.5281/zenodo.3519200}\relax
\mciteBstWouldAddEndPuncttrue
\mciteSetBstMidEndSepPunct{\mcitedefaultmidpunct}
{\mcitedefaultendpunct}{\mcitedefaultseppunct}\relax
\EndOfBibitem
\bibitem{supplemental}
{See Supplemental material to this Letter for further details.}\relax
\mciteBstWouldAddEndPunctfalse
\mciteSetBstMidEndSepPunct{\mcitedefaultmidpunct}
{}{\mcitedefaultseppunct}\relax
\EndOfBibitem
\end{mcitethebibliography}

\clearpage

\centerline
{\large\bf LHCb collaboration}
\begin
{flushleft}
\small
R.~Aaij$^{38}$\lhcborcid{0000-0003-0533-1952},
A.S.W.~Abdelmotteleb$^{57}$\lhcborcid{0000-0001-7905-0542},
C.~Abellan~Beteta$^{51}$\lhcborcid{0009-0009-0869-6798},
F.~Abudin{\'e}n$^{57}$\lhcborcid{0000-0002-6737-3528},
T.~Ackernley$^{61}$\lhcborcid{0000-0002-5951-3498},
A. A. ~Adefisoye$^{69}$\lhcborcid{0000-0003-2448-1550},
B.~Adeva$^{47}$\lhcborcid{0000-0001-9756-3712},
M.~Adinolfi$^{55}$\lhcborcid{0000-0002-1326-1264},
P.~Adlarson$^{84}$\lhcborcid{0000-0001-6280-3851},
C.~Agapopoulou$^{14}$\lhcborcid{0000-0002-2368-0147}
% ...

{\footnotesize \it

% ...
$^{14}$Universit{\'e} Paris-Saclay, CNRS/IN2P3, IJCLab, Orsay, France\\
% ...
$^{38}$Nikhef National Institute for Subatomic Physics, Amsterdam, Netherlands\\
% ...
$^{47}$Instituto Galego de F{\'\i}sica de Altas Enerx{\'\i}as (IGFAE), Universidade de Santiago de Compostela, Santiago de Compostela, Spain\\
% ...
$^{51}$Physik-Institut, Universit{\"a}t Z{\"u}rich, Z{\"u}rich, Switzerland\\
% ...
$^{55}$H.H. Wills Physics Laboratory, University of Bristol, Bristol, United Kingdom\\
% ...
$^{57}$Department of Physics, University of Warwick, Coventry, United Kingdom\\
% ...
$^{60}$School of Physics and Astronomy, University of Glasgow, Glasgow, United Kingdom\\
$^{61}$Oliver Lodge Laboratory, University of Liverpool, Liverpool, United Kingdom\\
% ...
$^{69}$Syracuse University, Syracuse, NY, United States\\
% ...
$^{84}$Department of Physics and Astronomy, Uppsala University, Uppsala, Sweden, associated to $^{60}$\\
% ...
}
\end{flushleft}

\end{document}